\documentclass[useAMS,usenatbib,12pt]{mn2e}
\usepackage[dvipdf]{graphicx,color}
\usepackage[latin1]{inputenc}
\usepackage{graphics}
\usepackage{amsfonts}
\usepackage{amsmath}
\usepackage{multicol}
\usepackage{layout}
\usepackage{amssymb}
\usepackage{epsfig}
\DeclareGraphicsExtensions{.eps}

\title[Halo Pairs]{The Cosmic Evolution of Halo Pairs: I. Global Trends}
\author[J. Moreno]{Jorge Moreno$^{1}$\\ \\
$^{1}$SISSA, Astrophysics Sector, via Bonomea 265, 34136, Trieste, Italy
}

\begin{document}
\date{}
\pagerange{\pageref{firstpage}--
\pageref{lastpage}} \pubyear{2011}
\maketitle
\label{firstpage}


\begin{abstract}

Accumulating evidence suggests that galaxy interactions play an important role in shaping the properties of galaxies.  For this reason, cosmological studies focused on the evolution of halo/subhalo pairs are vital.  In this paper I describe a large catalogue of halo pairs extracted from the publicly available Millennium Simulation \citep{springel05}, the largest of its kind to date.  (Throughout this work I use the term `halo' to refer both to individual haloes in the field {\it and} subhaloes embedded in larger structures.)  Pairs are selected according to whether or not they come within a given critical (comoving) distance $d_{\rm crit}$, {\it without} the prerequisite that they must merge.  Also, a condition requiring haloes to surpass a critical mass $M_{\rm crit}$ during their history is imposed.   The primary catalogue, consisting of 502,705 pairs, is selected by setting $d_{\rm crit}=1$ Mpc $h^{-1}$ and $M_{\rm crit} = 8.6 \times 10^{10} M_{\odot} h^{-1}$ (equivalent to 100 simulation particles).  One of the central goals of this paper is to evaluate the effects of modifying these criteria.  For this purpose, additional sub-catalogues with more stringent proximity and mass conditions are constructed (i.e., $d_{\rm crit}=200$ kpc $h^{-1}$ or/and  $M_{\rm crit}=8.6 \times 10^{11} M_{\odot} h^{-1}=1,000$ simulation particles --  see Table~\ref{setsTable} for a summary).  I use a simple {\it five-stage} picture to perform statistical analyses of their separations, redshifts, masses, mass ratios, and relevant lifetimes.  The fraction of pairs that {\it never} merge (because one of the members in the pair is absorbed by an third halo or both members survive until the present time) is accounted for. These results provide a broad picture that captures the essential characteristics behind the evolution of these halo pairs.  This is the first of a series of papers aimed to explore the huge wealth of information encoded in this catalogue.  Such investigations will play a fundamental role in future cosmological studies of interacting galaxies and binary (and multiple) quasars.  

\end{abstract}


\section{Introduction}\label{secIntro}

Galaxy mergers and interactions are among the most spectacular events in the sky.  But more importantly, they provide a bridge between worlds:  linking the large-scale hierarchical universe to the internal properties of individual galaxies and their evolution.  Such encounters have the ability to change the morphology of a galaxy \citep{toomre72}, and can cause a substantial migration of its gas towards the central regions \citep[e.g.,][]{barnes91,barnes96}.  Ultimately, sudden bursts of star formation and the activation of the supermassive black hole at the nucleus can be attributed to this mechanism. This single idea has been, without a doubt, instrumental to all modern theories of galaxy formation \citep[see e.g, the textbook by][and references therein]{mo10}.

Admittedly, mergers have received more attention than the early stages of interaction preceding them.  The last few years have witnessed the development of a whole industry of pre-prepared galaxy-galaxy merger simulations \citep[e.g.,][]{hopkins05,robertson06}.  Unfortunately, the pre-merger stages are seldom the priority of these numerical studies \citep[but see][whose simulations are tailored to emphasize the role of early interactions on star formation and black hole growth respectively]{dimatteo07,callegari09,callegari11}.  Likewise, huge efforts have also been reserved towards cosmological theoretical merger rates of galaxies, dark matter haloes and subhaloes \citep[e.g.,][]{lacey93,lacey94,governato99,gottlober01,cohn01,benson08,guo08,fakhouri08,neistein08,mateus08,wetzel09,genel09,moreno10,hester10,hopkins10}. 

Recently, however, a wide array of observations have provided hints that persuade us to redirect some of our attention to galaxy interactions.  For example, a number of studies demonstrate that star formation is enhanced in pairs (typically within $\lesssim 50$ kpc), and this effect is anti-correlated with their separation  \citep[e.g.,][and subsequent works by these groups]{barton00,lambas03,nikolic04,ellison08,darg10,wong11}.  Even at high redshifts, Ly$\alpha$ emission is heightened in pairs of Lyman-break galaxies (LBGs) with small separations \citep{cooke10}.

Galaxy interactions also play a role in active galactic nuclei (AGN).  For example, \cite{alonso07} report enhanced nuclear activity in $\sim$10\% of their galaxy pair catalogue (with projected separations $<25$ kpc $h^{-1}$).  The intensity of nuclear activity may depend on various factors, including the budget of gas available and the strength of the torques that can drive this gas into the central regions (the latter effect is primarily determined by the orbit of the encounter and the mass ratio).  Interestingly enough, they find that galaxies with bright/massive companions appear to be more active, suggesting that galaxies with shallower potentials are more susceptible to the gravitational torques produced during the interaction.




Lately, the search of binary quasars has gained impetus greatly in part because these systems are thought to be the precursors of coalescing black holes.  While it is true that supermassive black hole growth is expected to be dominated by gas accretion during quasar phases, not by mergers of black hole pairs \citep{salucci99,marconi04,shankar04,shankar09}; these unique events remain interesting, especially because their existence could be detected by future gravity-wave missions like the {\it Laser Interferometer Space Antenna} (LISA) \citep[e.g.,][]{sesana05}.  From a theoretical perspective, pairs of quasars at galactic separations establish the initial conditions necessary to comprehend these peculiar duos as they find their way into the central regions and perhaps into each other.

In particular, building up on previous samples \citep{hennawi06,myers07,myers08}, Foreman, Volonteri \& Dotti (2009) compiled a list of 85 quasar binary pairs.  Their physical-separation distribution peaks at $\sim$30 kpc, with a long tail at larger separations ($\sim$ 100-200 kpc).  Indeed, these authors find 14 pairs separated by distances $\gtrsim$ 300 kpc!  The size of this census was recently increased by over an order of magnitude by \cite{liu11b}, who report 1,288 binary AGNs (256 with tidal features), 39 triplets, 2 quadruplets and 1 quintuplet.  For the binaries, their projected-separation distribution exhibits peaks at $\sim$5, 20 and 60 kpc $h^{-1}$.  However, if only pairs displaying tidal features are considered, a sharper peak at $\sim$20 kpc $h^{-1}$ remains (at this stage, these are the only ones guaranteed to be truly interacting, and further investigations might be desirable for the rest).  Additionally, these authors find evidence that the galaxies in the pairs with the smaller stellar mass manifest more nuclear activity than their companions, suggesting that either secondary galaxies have more available fuel or that they are more susceptible to tidal perturbations \citep{liu11c}.  Lastly, supermassive black hole pairs have also been observed in dust-enshrouded merging galaxies.  For instance, \cite{dasyra06} find 21 ultraluminous infrared galaxies (ULIRGs) with nuclear separations between 1.6 and 23.3 kpc, thought to be between the first and final passage prior to a merger.

All this emerging observational evidence highlights the immediate need to focus on the long-lived evolution of galaxy pairs (and their supermassive black holes) in the stages {\it prior} to mergers.  With this ultimate goal in mind, I analyse a very large catalogue of halo pairs and their evolution, suitable for future statistical studies of galaxy and quasar binaries.  This catalogue, consisting of over half a million pairs, is extracted from the publicly available Millennium Simulation Database \citep{springel05}, the largest of its kind published so far.  In this work, pairs of haloes are selected by whether or not they are identified within a critical separation at some stage of their history.  With this choice, pairs of haloes that {\it never} merge are not automatically rejected by construction.  The effects of changing this critical condition are considered.  I also explore the consequences of selecting pairs involving haloes that reach masses above a given threshold at some point of their evolution.  The motivation here is to test whether close pairs of {\it massive} haloes are more or less susceptible to merging than pairs involving haloes with small masses.

This is the first of a series of papers that exploit this catalogue with the aim to understand the evolution of halo pairs in the universe.  As in any exploratory paper, I only touch on the most basic general trends.  These results will serve as a benchmark for future studies -- with more specialised sub-catalogues, refined recipes, and astrophysical applications -- to be presented in forthcoming papers.  Broadly speaking, I describe the evolution pairs in terms of five simple stages: {\it initial}, {\it entry}, {\it closest}, {\it final} and {\it fate} (see Section~\ref{subsecStages} below for definitions).  Quantities like the relative distance, redshift, total mass, mass ratio and lifetime are analysed statistically.  Kinematic vector quantities like velocities and spins will be analysed in a companion paper (Moreno 2011b, in preparation).  Relevant time intervals are also studied.  
Often throughout the paper I use the term `interacting' to refer to pairs within a chosen critical separation.  It is important to emphasize that I do not presume to know a priori the exact proximity criteria at which tidal disturbances are actually effective.  Nevertheless, using this term facilitates the discussion of the trends in the halo-pair catalogue.  

The remainder of this paper is organised as follows.  Section~\ref{secMethods} describes the methods: the Millennium Simulation, halo-pair identification, the catalogue and the stages of evolution.  Statistical results are found in Section~\ref{secResults}, which focus the stages and relevant lifetimes.  In Section~\ref{secDiscussion} I discuss the results, their interpretation, and their relation to other works in the literature. A summary is provided in Section~\ref{secSummary}.  Throughout this work, all distances are comoving, unless stated otherwise.

\section{METHODS}\label{secMethods}

\subsection{The Millennium Simulation}\label{subsecMS}

In a nutshell, the Millennium Simulation is a dark-matter only simulation that adopts a flat $\Lambda$CDM cosmology with the following parameters: $(\Omega_{\rm m},\Omega_{\rm b},\sigma_{8},n_{\rm s},h)=(0.25,0.045,0.9,1,0.73)$.  
The simulation covers a periodic comoving box with $L_{\rm box}=500$ Mpc $h^{-1}$ on the side.  It follows the evolution of $2160^3 \simeq 10^{10}$ dark matter particles with mass $m_{\rm p}=8.6 \times 10^{8} M_{\odot} h^{-1}$.  Dark matter haloes are identified using a friends-of-friends (\textsc{fof}) algorithm \citep{davis85} with a linking length set equal to 0.2 times the mean inter-particle separation.  These structures are identified at 64 snapshots from $z=127$ to the present, spaced nearly equally in $\log(1+z)$.  The Plummer-equivalent force softening scale is set to 5 kpc $h^{-1}$.

\begin{figure}
 \centering
 \includegraphics[width=\hsize]{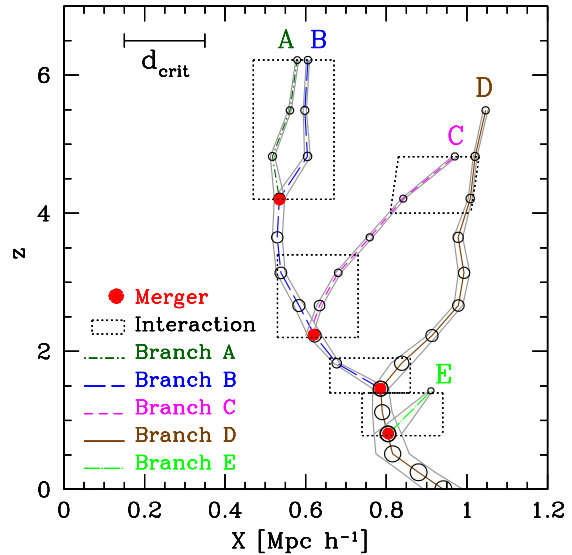}
 \caption{Interactions in a merger history tree:  redshift versus location (depicted schematically by the X-coordinate).  Haloes identified in the simulation are indicated by the open black circles, with radii proportional to the logarithm of the mass.  Mergers are represented by solid red circles.  Interactions are enclosed in dotted black parallelograms.  Five branches are shown, with colours and line-styles indicated.  The solid gray lines surrounding the branches enclose half the mass.  Of the 10 potential branch pairs, only five pairs interact within $d_{\rm crit}=$ 200 kpc $h^{-1}$.   Of those, all result in a merger, except for the C-D pair.  Branch E is tidally-stripped below resolution before merging on to branch D (an open circle on branch E at $z \simeq 1.2$ is missing).}
 \label{figTree}
\end{figure}

Self-bound structures with the \textsc{fof} groups are identified with the \textsc{subfind} algorithm \citep{springel01}.  This method allows to identify the main background dark matter halo and its substructures (i.e., the subhaloes).  \cite{giocoli10} compares \textsc{subfind} against various other substructure-finding algorithms in their Appendix~A.  Merger history trees are built from these self-bound objects by tracking their `descendants' at the next snapshot with lower redshift \citep[see the supplementary information in][for more details]{springel05}.   Once the trees are built, one can track the identity of each object in the hierarchy back in time by following its {\it principal branch} (see Figure~\ref{figTree}, and its associated discussion below).  This is roughly done by listing the most-massive progenitor, its subsequent most-massive progenitor, and so forth.  See \cite{delucia07} for a discussion of the subtleties in assigning the most-massive progenitor to an object and their improved `first progenitor' designation  (hereafter I drop the word `principal' unless it is absolutely necessary).  

A few comments on terminology are in order.  In this paper, the words `subhalo' or `satellite' always refer to a self-bound object immersed in another larger self-bound structure (the latter is often called the `background halo' or the `host halo').  Objects that are not substructures of other objects are called `isolated' haloes.  For instances, subhaloes were isolated haloes before becoming members of a larger halo.  By definition, host haloes are isolated haloes with substructure.  In this work the term `{\it halo}' refers to any self-bound structures identified by the the \textsc{subfind} algorithm, without regard to whether they live in a host background halo or not. An upcoming paper will exploit these environmental differences.  

Lastly, defining {\it mergers} is not entirely trivial in N-body simulations because a subhalo may go below resolution before having a chance to merge with another structure.  Many authors adopt analytic dynamical-friction recipes to track down mergers \citep[see e.g.,][and references therein]{kitzbichler08}.  For the sake of simplicity, if two haloes share a descendant at the next snapshot, they are considered as merging.  More sophisticated dynamical-friction techniques can always be incorporated in future applications of this catalogue.  See Section~\ref{secDiscussion} for more comments on this issue.


\begin{figure}
 \centering
 \includegraphics[width=\hsize]{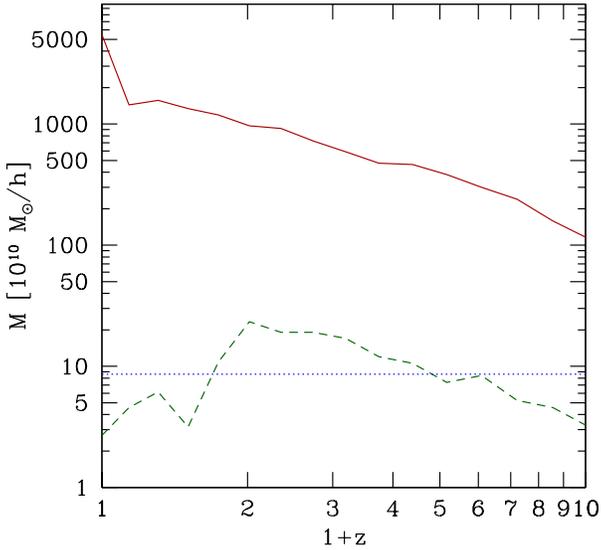}
 \caption{The mass growth of two interacting haloes.  The more massive of the two begins at $\simeq 1.2 \times 10^{12} M_{\odot} h^{-1}$ and grows up to $\simeq 5 \times 10^{13} M_{\odot} h^{-1}$ (solid light-red line).  The secondary halo begins at $\simeq 3 \times 10^{10} M_{\odot} h^{-1}$, reaches a maximum mass of $\simeq 3 \times 10^{11} M_{\odot} h^{-1}$ (at $z \simeq 1$), and finishes at $z=0$ with $\simeq 2.5 \times 10^{10} M_{\odot} h^{-1}$ (dashed dark-green line).  The horizontal blue dotted lines represents the threshold that selects which branches make it to the catalog (set at 100 particles $=8.6 \times 10^{10} M_{\odot} h^{-1}$). Branches are selected by whether they cross this threshold at some point in their history, regardless of their final mass.}
 \label{figThreshold}
\end{figure}

\subsection{Identifying the Halo Pairs}\label{subsecIdentifying}

The goal of this Section is to establish the language necessary to describe the evolution of close pairs of haloes.  To do this, I discuss a single merger history tree extracted from the Millennium Simulation in detail.  In reality, the spatial evolution of a merger tree should be represented in $(3+1)$-dimensions.  However, for illustrative purposes, the $(1+1)$-dimensional representation in Figure~\ref{figTree} suffices.

In this Figure, the black open circles depict the redshifts and positions (schematically represented by the X-coordinates) of the haloes identified in the simulation.  The radii are proportional to the logarithm of the mass.  The solid gray lines flanking the branches enclose half of the mass of each halo, giving a rough sense of the `size' of these objects.  The solid red circles denote mergers and the black dotted rectangles (or parallelograms) represent close interactions.  The branches are said to be `interacting' if they are within some critical distance $d_{\rm crit}$ of each other (in the Figure, $d_{\rm crit} = $ 200 kpc $h^{-1}$).  

The merger history occurs as follows.  First branch A (dark-green dot-dashed lines) merges with branch B (blue long-dashed lines).  Of these two, only branch B continues on to later epochs after the merger.  Another way to put this is that branch A `merges on to' branch B because, right before the merger, the halo on branch B is more massive than the halo on branch A.  Later on, branch C (magenta short-dashed lines) also merges on to branch B.  This is followed by branch B merging on to branch D (dark-brown solid lines). Lastly, branch E (dotted long-dashed green lines) merges on to branch D.    In this merger tree, branch D is the principal branch of the final halo rooted at $z=0$.

Not all branches begin at the same redshift. For instance: branches A and B appear simultaneously. Later, branch D appears, followed by branch C.  Branch E is born long after most of the other branches have already merged.  Physically, the haloes might have been around from much earlier epochs.  However, the simulation only identifies haloes with at least 20 particles.  

The growth of haloes within larger haloes (i.e., subhaloes) is commonly thwarted by the rich environments surrounding them, which tidally strip them of their diffuse outer layers \citep[e.g.,][]{moore96,moore99,tormen97,tormen98,ghigna98,klypin99,taffoni03,giocoli08}.  In many cases, haloes go below the resolution of the simulation, and are no longer identified.  Such is the case of branch E, which disappears before merging on to branch D. Extra care must be taken when dealing with these situations.  Namely, one must pay attention and make sure that both haloes are identified at every snapshot in the interaction. 
In this work I impose an extra condition on the mass which deletes many (but not all) of the small sub-haloes that get stripped apart by their hosts (see Section~\ref{subsubsecMassSelection} below).  With this condition, much less than one percent of the pairs have one member going below resolution (situations where both members go below resolution are non-existent under this condition).  For the remainder of the paper, these effects will be ignored because they happen so infrequently.  

A merger tree with $N$ branches has $N(N-1)/2$ candidate pairs.  However, not all of them qualify as interacting.  For instance, the merger tree in Figure~\ref{figTree} has 5 branches and $5(5-1)/2=10$ candidate pairs; but 5 of those 10 pairs {\it never} count as `close' pairs.  Namely, the distance between branch A and either branches C or D is always greater than $d_{\rm crit}$.  Moreover, branch E appears after branches A, B and C have already merged on to branch D.  In other words, branch E never `coexists' with branches A, B and C.  Although there are five closely-interacting pairs, only four end up as mergers.  Pair C-D interacts briefly but eventually gets separated beyond $d_{\rm crit}$.  Branch C never has the opportunity to merge on to branch D, and instead merges on to branch B.  It is often tempting to assume that all close pairs merge eventually.  One of the goals of this paper is to quantify the fraction of close pairs that {\it never} merge.

Figure~\ref{figTree} only shows a single merger tree.  However, many massive haloes often have plenty of substructures.  The appropriate thing to do in such situations is to follow the merger histories of both the central dominant halo {\it and} the satellites.  In this paper, I pair up all branches in a given group, even if they belong to different merger trees within that group.
Often I will use the term `never-merging' to refer to pairs that do not merge because either one of the members in the pair gets absorbed by a third halo or because both halos survive as separate substructures of a larger background halo until the present time.

 Lastly, one could also pair up branches belonging to different groups; i.e., pairs of haloes that never end up gravitationally bound to each other.   These events could be important if one tries to use close pairs as a proxy of mergers \citep[e.g.,][]{wetzel08}, especially if large separations (of the order of 1 Mpc $h^{-1}$) are used.  This contribution is more likely to be at the few-percent level if smaller separations (of the order of 200 kpc $h^{-1}$) are used (private communication with the reviewer).  For the sake of simplicity, in this work I ignore those pairs whose haloes end up in different host haloes.  The effects of this particular limitation will be addressed in future work in the context of close-pair counts.

\subsection{The Stages of Evolution}\label{subsecStages}

In this paper, the evolution of a pair of closely interacting haloes is described in terms of the following {\it five} stages: 

\begin{itemize}

\item {\bf Initial:}  When the two haloes both exist for the first time (above the resolution of the simulation), regardless of whether they are closely interacting or not.

\item {\bf Entry:}  When the two haloes are, for the first time, at a distance less than $d_{\rm crit}$.  This may occur either at the initial stage or later.

\item {\bf Closest:}  When the two haloes are both identified as separate entities by the simulation and are at their closest distance.

\item {\bf Final:}   The last time the two haloes are both identified by the simulation, regardless of whether they are closely interacting or not. 

\item {\bf Fate:}   What ultimately happens to the pair: either (1) they merge, (2) one or both haloes are absorbed by a third halo or (3) both haloes survive until today.

\end{itemize}

It is useful to clarify a few subtleties in these definitions with specific examples from Figure~\ref{figTree}.  First of all, notice that both haloes in the A-B pair appear simultaneously and are both very small (just above resolution, with mass ratio very close to one) at the \underline{initial} time.  In other cases, however, one of the two haloes may have some time to grow before its partner shows up in the simulation (pairs B-C and C-D, and especially D-E -- the latter with a mass ratio particularly different from unity).  

When the initial distance of the pair is below $d_{\rm crit}$, the \underline{entry} and initial times coincide (pairs A-B, C-D and D-E).  For pairs A-C and A-D, on the other hand, the initial distance is larger than $d_{\rm crit}$; and the entry time occurs much later than the initial time.  When the entry time does not correspond to one of the snapshots in the simulation (e.g., as in pairs B-C and B-D), a linear interpolation in $\log(1+z)$ is performed.

The \underline{closest} distance is often equal to the final distance, especially right before mergers (pairs B-C, B-D and D-E).  In some cases, though, the closest distance does not match the final distance.  Examples of these include pairs A-B and C-D; where the former results in a merger and the latter does not.  

The \underline{final} epoch refers to the last time both haloes are identified in the simulation as separate entities.  They need not be within $d_{\rm crit}$: for instance the final time for pair C-D takes place at $z\simeq2.7$, with separation $\simeq 2 d_{\rm crit}$.  In this particular case, it is interesting to know when the pair is within $d_{\rm crit}$ for the last time (i.e., $z\simeq4$ for pair C-D).  In Section~\ref{subsecResultsDuration}, I use this information to estimate the typical `lifetime' of interactions.

The ultimate \underline{fate} stage is formally {\it not} considered part of the evolution of the pair.   Failing to make this distinction may lead to over-counting the number of instances a halo participates in an interaction. For example, the halo resulting from the merger of branches A and B participates in both pairs B-C and B-D, but not in A-B itself. (Once the merger has happened, the single remnant halo cannot be counted as part of its own halo pair!)  It is important to probe the interval between the final and the fate time.  If this portion is not included towards the lifetime of the pair, one could conclude that pair D-E has zero duration, whereas in reality it lasts from $z\simeq 2.4$ down to $z\simeq 1.8$ (slightly less than three full snapshots).  
Nevertheless, as mentioned before, in this paper I impose a mass condition that deletes most pairs involving haloes going below resolution (see section~\ref{subsubsecMassSelection} ) -- thus situations like the D-E pair are extremely uncommon with this condition.  Lastly, if the final epoch takes place today, the fate epoch is set equal to the final epoch.  This is the case for pairs involving different members of a group today.

\subsection{Constructing the Catalogue}\label{subsecCatalogue}

In this Section I explain the construction of the halo pair catalogue used in this paper, which can be understood as a series of selection criteria.  I defer discussions on the adopted choices and assumptions to Section~\ref{subsecDiscussionCatalogue}.

\subsubsection{Mass Selection}\label{subsubsecMassSelection}

First, I only consider branches that reach a critical mass $M_{\rm crit}$ at least once at some point throughout their lifetime.  Isolated haloes typically reach their maximum mass at $z=0$.  However, recall that a halo immersed in a larger group (i.e., a subhalo) may be substantially dissolved by the gravitational tidal field of its host.  In this case, it is possible that at some point a given halo had mass $>M_{\rm crit}$, but ends up with mass $<M_{\rm crit}$.  In some extreme cases, it may even go below the resolution of the simulation.  It must be mentioned that for general branches (i.e., those not obeying the above mass condition), this situation are not uncommon.  Nevertheless, once the above mass condition is imposed, such situations become very infrequent (less than one percent of the pairs have one member going below resolution at the late stages of interaction).

In this work, massive branches are kept even if they lose a substantial fraction of their mass at later epochs.  To select these branches, the appropriate thing to do is to choose them by their maximum mass, {\it not} their final mass.  Figure~\ref{figThreshold} illustrates this situation by showing the mass history of two interacting haloes that are both accepted into the catalogue by this criterion.  The horizontal dotted blue line denotes the mass selection threshold (in the Figure, $M_{\rm crit}=8.6\times10^{10} M_{\odot} h^{-1}=100 \times m_{\rm p}$). The solid light-red line represents the more massive halo, while the secondary halo is shown as a dark-green dashed line. The first branch grows from $\simeq 1.2 \times 10^{12} M_{\odot} h^{-1}$ up to $\simeq 5 \times 10^{13} M_{\odot} h^{-1}$.  For the most part, the mass of this halo increases monotonically (except for the brief dip near $z \simeq 0.2$).  At all times this branch is above the selection threshold.  The second branch begins at $\simeq 3 \times 10^{10} M_{\odot} h^{-1}$, below the selection threshold.  At $z\simeq 3.5$, it crosses the threshold -- reaching a maximum mass of $\simeq 3 \times 10^{11} M_{\odot} h^{-1}$ at $z\simeq 1$.  At that point, it begins to be tidally stripped of its mass and goes below the selection threshold at $z\simeq 0.7$.  By $z=0$, it ends up with a final mass of $\simeq 2.5 \times 10^{11} M_{\odot} h^{-1}$.   If the selection were based on the mass today, this branch would have been thrown out, despite the fact that at some point it had a mass four times greater than the threshold!

In summary, only branches that reach a critical mass $M_{\rm crit}$ at some point in their existence are paired up in the catalogue.  This maximum mass is usually attained at the end of their lives as isolated haloes, right before they are accreted by larger haloes.  This is often called the `infall mass', and many authors consider it an important ingredient in galaxy formation models \citep[e.g.,][]{conroy06,moster10}.  

\subsubsection{Co-existence}\label{subsubsecCoexistence}

With the list of branches satisfying the above mass selection criterion, the next step is to collect pairs.  A tree with $N$ such branches can have up to $N(N-1)/2$ pairs.  However, in many cases, a branch disappears (it merges on to another branch) long before the candidate partner shows up in the simulation (see discussion about branch E in regards to branches A, B and C; in Section~\ref{subsecIdentifying} and Figure~\ref{figTree}).
The correct way to check whether two branches `co-exist' is to see if the intervals from birth (when the branch first appears in the simulation) to fate of the two branches ever match.  Special care must be taken when branches go below resolution, typically towards the end of their existence.

\subsubsection{Distance Selection}\label{subsubsecDistanceSelection}

Once the above two criteria have been satisfied, one can readily identify the initial and final stages, as explained in Section~\ref{subsecStages}.  The relative distance can be calculated for all those snapshots for which both haloes are identified. With this, the next step is to find the minimum distance $d_{\rm min}$ (the closest stage, Section~\ref{subsecStages}), and check if the $d_{\rm min} < d_{\rm crit}$ criterion is met.

\subsubsection{The Halo-Pair Sets}\label{subsubsecSets}

The following sets of halo-pairs are considered: the {\bf Close Set} (C), the {\bf Very Close Set} (VC), the {\bf Massive Close Set} (MC) and the {\bf Massive Very Close Set} (MVC).   These sets are defined by the following critical parameters:

\begin{itemize}

\item    C:  $(d_{\rm crit},M_{\rm crit})=(1$ Mpc $h^{-1},8.6\times10^{10} M_{\odot} h^{-1})$.
\item VC:  $(d_{\rm crit},M_{\rm crit})=(200$ kpc $h^{-1},8.6\times10^{10} M_{\odot} h^{-1})$.
\item MC:  $(d_{\rm crit},M_{\rm crit})=(1$ Mpc $h^{-1},8.6\times10^{11} M_{\odot} h^{-1})$.
\item MVC:  $(d_{\rm crit},M_{\rm crit})=(200$ Mpc $h^{-1},8.6\times10^{11} M_{\odot} h^{-1})$.

\end{itemize}
Notice that the latter three sets are simply sub-catalogues of set C.  Often in the text, and in all the remaining Figures, I use the following terminology:  C and MC are the {\bf Close Sets} (with $d_{\rm crit}=1$ Mpc $h^{-1}$, upper panels);  VC and MVC are the {\bf Very Close Sets} (with $d_{\rm crit}=200$ kpc $h^{-1}$, lower panels); C and VC are the {\bf non-Massive Sets} (with $M_{\rm crit}=8.6\times10^{10} M_{\odot} h^{-1}$, left-hand panels); and MC and MVC are the {\bf Massive Sets} (with $M_{\rm crit}=8.6\times10^{11} M_{\odot} h^{-1}$, right-hand panels). For quick reference, Table~\ref{setsTable} indicates the assigned values of $d_{\rm crit}$ and $M_{\rm crit}$, and the above terminology.


\begin{figure*}
 \centering
 \includegraphics[width=\hsize]{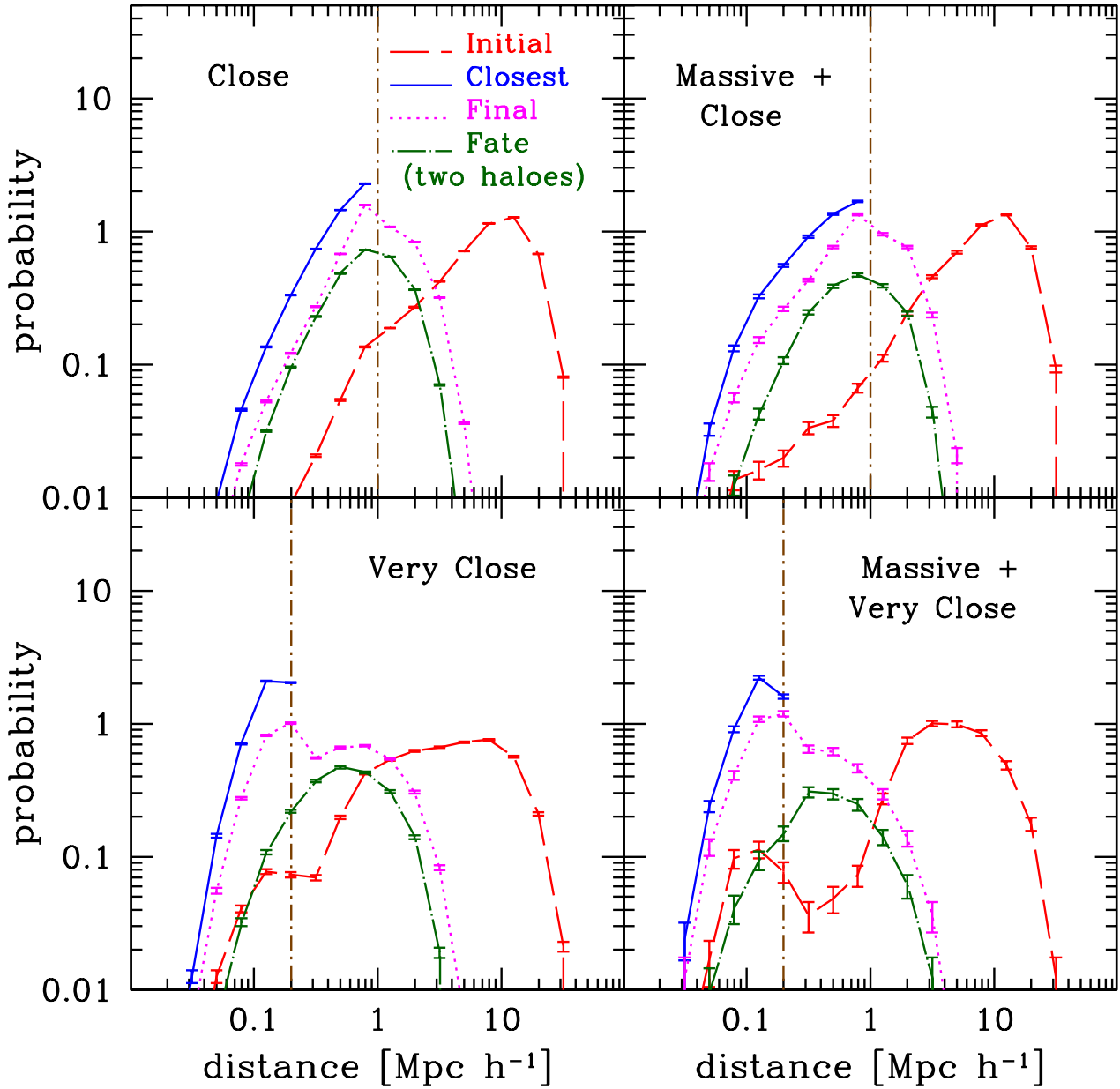}
 \caption{The relative-distance distribution of halo pairs.  The set of halo pairs presented (see Table~\ref{setsTable}) is indicated in each panel: C (upper-left), VC (lower-left), MC (upper-right) and MVC (lower-right).  The colours and line-styles are indicated in the Figure. The vertical line represents $d_{\rm crit}$.  Only pairs that never merger are presented for the fate stage (indicated by the `two haloes' label). For reference, the text uses the following definitions to describe the sets:  Close Sets ($d_{\rm crit}=1$ Mpc $h^{-1}$; upper panels), Very Close Sets ($d_{\rm crit}= 200$ kpc $h^{-1}$; lower panels), non-Massive Sets ($M_{\rm crit}=8.6 \times 10^{10} M_{\odot} h^{-1}$; left-hand panels), and Massive Sets ($M_{\rm crit}=8.6 \times 10^{11} M_{\odot} h^{-1}$; right-hand panels).  See Section~\ref{subsecStages} and \ref{subsubsecSets} respectively for definitions and descriptions of the stages and sets.}
 \label{figDistancePanel}
\end{figure*}

\begin{center}
\begin{table}
\begin{tabular}{| c | c | c | c |l c}
\hline
Set & $d_{\rm crit}$ & $M_{\rm crit}/M_{\odot}h^{-1}$ & $M_{\rm crit}/m_{\rm p}$ & Number\\
\hline \hline
{\bf C}  & $1$ Mpc $h^{-1}$ & $8.6 \times 10^{10}$ & $100$ & 502,705\\
{\bf VC} & $200$ kpc $h^{-1}$ & $8.6 \times 10^{10}$ & $100$ & 32,582 \\
{\bf MC} & $1$ Mpc $h^{-1}$ & $8.6 \times 10^{11}$ & $1,000$  & 13,637 \\
{\bf MVC} & $200$ kpc $h^{-1}$ & $8.6 \times 10^{11}$ & $1,000$ & 2,067 \\
\hline
\end{tabular}
\caption{Halo pair sets shown in the panels of all Figures in Section~\ref{secResults}.  The abbreviations are:  C ({\it Close}), VC ({\it Very Close}), MC ({\it Massive Close}) and MVC ({\it Massive Very Close}).   Distances are in comoving coordinates.  Recall that $m_{\rm p} = 8.6 \times 10^{8} M_{\odot} h^{-1}$.  For reference, C and MC are the {\it Close Sets}, VC and MVC are the {\it Very Close Sets}, C and VC are the {\it non-Massive Sets}, and MC and MVC are the {\it Massive Sets}.}
\label{setsTable}
\end{table}
\end{center}

\begin{figure*}
 \centering
 \includegraphics[width=\hsize]{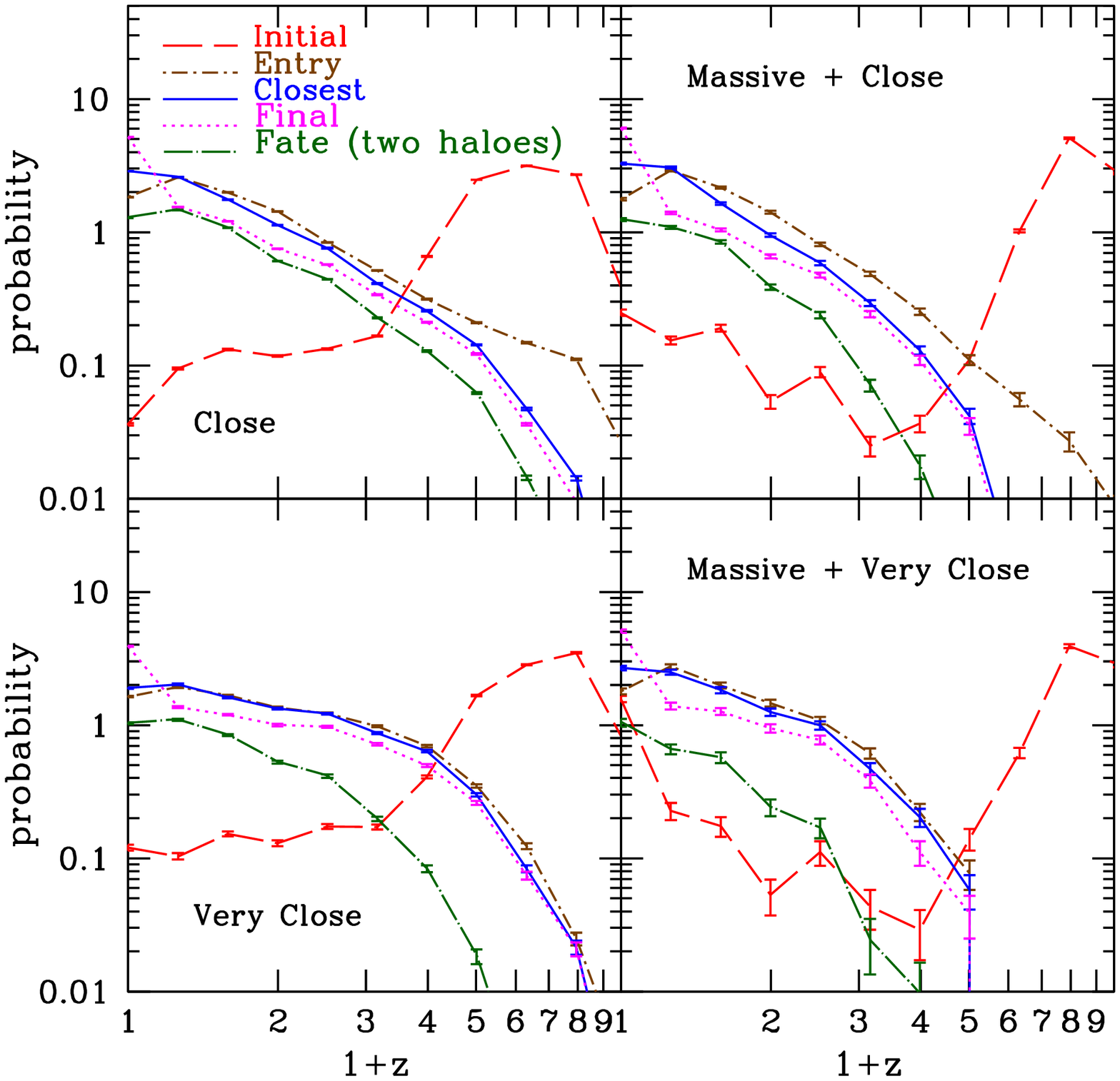}
 \caption{The redshift distribution of halo pairs.  The set of halo pairs presented (see Table~\ref{setsTable}) is indicated in each panel: C (upper-left), VC (lower-left), MC (upper-right) and MVC (lower-right).  The colours and line-styles are indicated in the Figure.  Only pairs that never merger are presented for the fate stage (indicated by the `two haloes' label). For reference, the text uses the following definitions to describe the sets:  Close Sets ($d_{\rm crit}=1$ Mpc $h^{-1}$; upper panels), Very Close Sets ($d_{\rm crit}= 200$ kpc $h^{-1}$; lower panels), non-Massive Sets ($M_{\rm crit}=8.6 \times 10^{10} M_{\odot} h^{-1}$; left-hand panels), and Massive Sets ($M_{\rm crit}=8.6 \times 10^{11} M_{\odot} h^{-1}$; right-hand panels).  See Section~\ref{subsecStages} and \ref{subsubsecSets} respectively for definitions and descriptions of the stages and sets.}
 \label{figRedshiftPanel}
\end{figure*}

\section{RESULTS}\label{secResults}

In this Section I present the distribution of several quantities as a function of what stage in their evolution the pairs happen to be.  I also address relevant lifetimes and the effects of modifying $d_{\rm crit}$ and $M_{\rm crit}$.  Only results are reported here. I postpone associated interpretations and discussions for Section~\ref{secDiscussion}.  I do this for the sake of clarity because the forthcoming discussion requires having described all the Figures in this Section already.


\subsection{Relative Distances}\label{subsecDistancesPanel}

Each panel in Figure~\ref{figDistancePanel} shows each of the four sets in Table~\ref{setsTable}. The vertical line is drawn at $d_{\rm crit}$ to indicate the separation at the {\it entry} stage.  Relevant colours and line-styles are indicated in all Figures.

Typically, pairs have large \underline{initial} distances, between a few up to $10$ Mpc $h^{-1}$, but never beyond $30$ Mpc $h^{-1}$.  There is, however, a small fraction of pairs ($\sim$5\%) with $d_{\rm initial}<d_{\rm crit}$.  The most common \underline{closest} separation is near $d_{\rm crit}$. Recall that close branches are selected so $d_{\rm closest} < d_{\rm crit}$, which is why this distribution is truncated at $d_{\rm crit}$.  At the \underline{final} stage, a large fraction of pairs end up at distances beyond $d_{\rm crit}$.  For the rest, the final-distance distribution resembles the closest-distance distribution.  In all cases, a non-negligible fraction of pairs {\it never} merge (labeled as `\underline{fate} with two haloes'), and many of those even end up at distances greater than $d_{\rm crit}$.  Below, I describe these trends for each stage separately.

\subsubsection{Initial Distances}\label{subsubsecInitialDistancesPanel}

The initial-distance distribution for the Close Sets peaks at $\simeq 10$ Mpc $h^{-1}$, is narrow, and drops sharply at low separations.   For the Very Close Sets, the distribution peaks around $8$ and $3$ Mpc $h^{-1}$ (for the VC and MVC sets respectively), is wider, and has a small secondary peak at $\simeq 100$ kpc $h^{-1}$ before dropping sharply at low separations (the secondary peak is more evident for the MVC set).  
Reducing $d_{\rm crit}$ does not alter the distributions significantly, except for the appearance of a bump slightly below $d_{\rm crit}$ and a trough slightly above this value.  Increasing $M_{\rm crit}$ also enhances the low-separation regime, but to a very small degree.  I must emphasize that the initial stage depends strongly on the resolution of the simulation.  Nevertheless, it is interesting to show the highest-redshift status of the close pairs one can study with the Millennium Simulation.



\subsubsection{Closest Distances}\label{subsubsecClosestDistancesPanel}

For the most part, the closest-distance distribution increases monotonically with increasing distance and is truncated at $d_{\rm crit}$.  For the Very Close Sets, the peak coincides with the location of the secondary peak in the initial-distance distribution. 
Reducing $d_{\rm crit}$ strongly increases the contribution from pairs with small minimum separations (in particular, below $100$ kpc $h^{-1}$).  Increasing $M_{\rm crit}$ leaves the distribution almost intact, except for a slight enhancement at small separations.

\subsubsection{Final Distances}\label{subsubsecFinalDistancesPanel}

The shape of final-distance distribution mimics that of the closest-distance distributions up to $d_{\rm crit}$, and then decreases beyond that (with an extra small bump at $\simeq 800$ kpc $h^{-1}$ for the Very Close Sets).  The final distance never exceedes $5$ Mpc $h^{-1}$. For the Close Sets, the final distance distribution peaks at $\sim800$ kpc $h^{-1}$, while the peak is at $\sim200$ kpc $h^{-1}$ for the Very Close Sets. 

Decreasing $d_{\rm crit}$ shifts the distribution to smaller separations, enhancing the number of small-distance pairs (with $d_{\rm final}<200$ kpc $h^{-1}$) quite significantly.  However, this change is not so strong for pairs with $d_{\rm final}<d_{\rm crit}$.  On the other hand, the contribution to other regimes ($>200$ kpc $h^{-1}$) is suppressed with decreasing $d_{\rm crit}$.  The effects of increasing $M_{\rm crit}$ are similar to those of decreasing $d_{\rm crit}$, but milder.  However, while decreasing $d_{\rm crit}$ enhances the secondary bump present in the Very Close Sets, increasing $M_{\rm crit}$ actually suppresses the bump, but only slightly.


\subsubsection{Fate Distances (Two Haloes)}\label{subsubsecFateDistancesPanel}

I only present the fraction of fate distances of pairs that {\it never} merge.  (Recall that never-merging pairs are those that either get split by a third party or survive as two entities until today.) 
For the fate two-halo distribution, I will often speak of the `absolute' and `relative' contribution from a given regime.  The former is normalised to the total population of pairs, while the latter is normalised to the subset of pairs that never merge.  In other words, the sum of all {\it relative} contributions is 100\%, while the sum of all {\it absolute} contributions is one of the four percentages quoted above.

The shape of the fate-distance distribution is similar to that of the final-distance distributions for the Close Sets, and to the secondary bump in the final-distance distributions for the Very Close Sets.  Reducing $d_{\rm crit}$ decreases the absolute number of pairs that never merge. 
It shifts the distribution to lower distances (the peak moves from $\sim800 \rightarrow \sim400$ kpc $h^{-1}$ for the non-Massive Sets, and from $\sim800 \rightarrow \sim300$ kpc $h^{-1}$ for the Massive Sets).   However, the relative fraction of pairs with $d_{\rm fate}>d_{\rm crit}$ increases.  In other words, there is a shift to smaller separations; but this shift is actually to larger separations in units of $d_{\rm crit}$.  Increasing $M_{\rm crit}$ also suppresses the absolute contribution of pairs that never merge. This effect is more evident than that produced by reducing $d_{\rm crit}$. 


\subsection{Redshifts}\label{subsecRedshiftsPanel}

Figure~\ref{figRedshiftPanel} shows the distributions of halo-pair redshifts at the various stages of their evolution.  
The \underline{initial}-redshift distribution is dominated by pairs at very early epochs ($z>4$), with some low-redshift contributions for the Massive Sets.  
The subsequent stages of interaction (i.e., \underline{entry} $\rightarrow$ \underline{closest} $\rightarrow$ \underline{final}) happen at much lower redshifts (compared to the initial distribution), and are increasingly less frequent with high redshift.  As the interaction progresses, the distribution shifts to lower redshifts (the final distribution exhibits a small upturn at very low redshifts).  This trend continues at the \underline{fate} stage for those pairs that never merge.  More than $\sim$50\% of the never-merging fate pairs occur at $z<0.5$, while less than $\sim$10\% take place at $z>1$.







Reducing $d_{\rm crit}$ only enhances the fraction of initial pairs below $z\sim1$ slightly. This change makes the entry-redshift distribution shallower in the $z<3$ regime, and introduces a sharp drop afterwards, bringing its shape more in line with the closest and final-redshift distributions.  The same flattening and drop appear for the closest and final distributions.  The fate distribution is suppressed more strongly as a function of redshift, increasing the discrepancy with the other three distributions described here.
Increasing $M_{\rm crit}$ also enhances the fraction of initial pairs below $z=1$.  Additionally the peak near $z\sim6$ becomes more narrow, and a trough at $z \sim 3$ appears.   For the entry distribution, the high-redshift regime is slightly suppressed (but not completely erased) for the Close Sets, and completely cut off beyond $z>4$ for the Very Close Sets.  The closest and final distributions are also cut-off completely beyond $z>4$.  The fate distributions are completely suppressed beyond $z>3$, and enhanced in the $z<0.5$ regime.

\begin{figure*}
 \centering
 \includegraphics[width=\hsize]{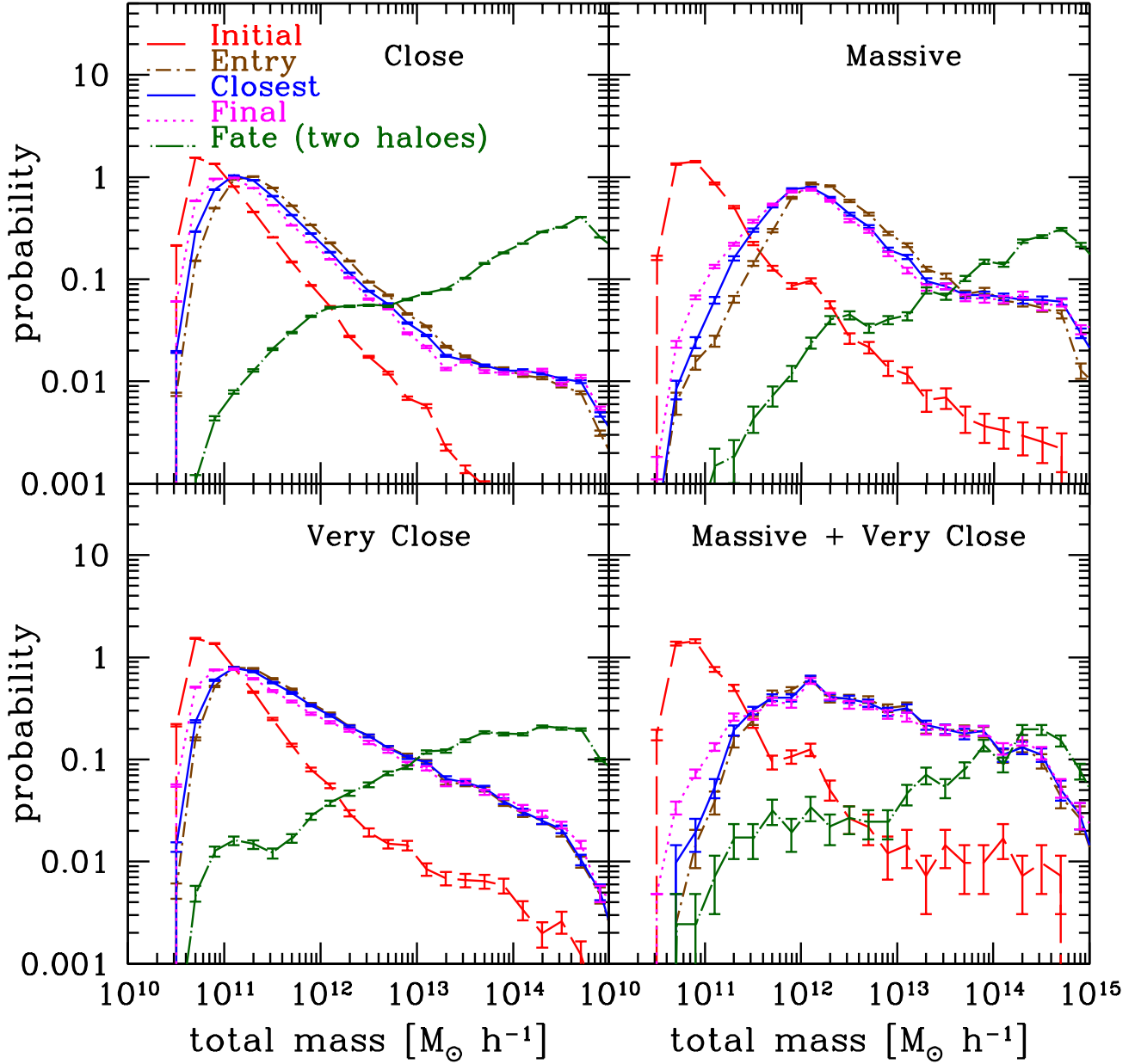}
 \caption{The total mass distribution of halo pairs.   The set of halo pairs presented (see Table~\ref{setsTable}) is indicated in each panel: C (upper-left), VC (lower-left), MC (upper-right) and MVC (lower-right).  The colours and line-styles are indicated in the Figure.  Only pairs that never merger are presented for the fate stage (indicated by the `two haloes' label). For reference, the text uses the following definitions to describe the sets:  Close Sets ($d_{\rm crit}=1$ Mpc $h^{-1}$; upper panels), Very Close Sets ($d_{\rm crit}= 200$ kpc $h^{-1}$; lower panels), non-Massive Sets ($M_{\rm crit}=8.6 \times 10^{10} M_{\odot} h^{-1}$; left-hand panels), and Massive Sets ($M_{\rm crit}=8.6 \times 10^{11} M_{\odot} h^{-1}$; right-hand panels).  See Section~\ref{subsecStages} and \ref{subsubsecSets} respectively for definitions and descriptions of the stages and sets.}
 \label{figMassPanel}
\end{figure*}

\subsection{Total Masses}\label{subsecTotalMassesPanel}

Figure~\ref{figMassPanel} shows the distribution of the sum of the two masses (hereafter, the {\it total mass}) in a halo-pair at various stages of their evolution.  
At the \underline{initial} stage, the majority of the total masses are rather small (peaking below $10^{11} M_{\odot} h^{-1}$, with a negligible contribution from masses greater than a few $\times 10^{13} M_{\odot} h^{-1}$).  The transition from the initial to the \underline{entry} stage populates the large-mass regime, evacuating the small-mass regime as a result.  This shift towards more massive pairs is more evident for the Massive Sets.
As the interaction progresses (i.e., \underline{entry}, \underline{closest} and \underline{final} stages), the very large-mass regime ($\gtrsim$ a few $\times 10^{13} M_{\odot} h^{-1}$) continues to grow, while the rest shrinks. (This transition mass scale is not very evident for the MVC set.)  These shifts are very small relative to the previous initial $\rightarrow$ entry shift, but not entirely insignificant.
The \underline{fate}-mass distribution of pairs that never merge does not resemble the others because it is controlled by third parties that absorb the pairs, not by the pairs themselves.  Notice that these pairs are incorporated preferentially into extremely massive haloes (ending up with masses typically greater than $10^{14} M_{\odot} h^{-1}$).  These situations most likely involve two subhaloes interacting for a while until they are split apart by the central halo.

\subsubsection{Initial Total Masses}\label{subsubsecInitialTotalMassPanel}

Reducing $d_{\rm crit}$ does not alter the initial-mass distribution by much, except for a small enhancement in the large-mass regime (with masses $\gtrsim$ a few $\times 10^{13} M_{\odot} h^{-1}$).  A similar effect happens when $M_{\rm crit}$ is increased.  This distribution is strongly determined by the resolution of the simulation, and its features will not be addressed in much detail.  Nevertheless, it serves as reference to see how much the branches grow during the interacting stages.

\subsubsection{Entry, Closest, and Final Total Masses}\label{subsubsecInteractionTotalMassPanel}

In this Section I describe the entry, closest and final stages together because they follow similar total-mass distributions.  These distributions are broader than the initial-mass distribution, and peak at a higher values: slightly above $10^{11}$ and $10^{12} M_{\odot} h^{-1}$ for the non-Massive and Massive Sets respectively.  Relative to the initial stage, the typical total mass increases by a factor of $\sim2$ for the non-Massive Sets, and by a factor of $\sim20$ for the Massive Sets.  Also, the large-mass contribution (with masses $\gtrsim$ a few $\times 10^{13} M_{\odot} h^{-1}$) is no longer negligible.  

Reducing $d_{\rm crit}$ does not alter the distributions much, except that the high-mass drop becomes shallower, enhancing the contribution from that regime.  This is particularly noticeable for the Massive Sets.
Increasing $M_{\rm crit}$ shifts the peak of the distributions from $ \simeq 10^{11} \rightarrow 10^{12} M_{\odot} h^{-1}$.  This suppresses the small-mass regime, and enhances the large-mass regime.  The location of the new peak coincides with that of the small bump found in the initial total-mass distribution.  The large-mass plateau is enhanced more significantly here than when $d_{\rm crit}$ is reduced.


\subsubsection{Fate Total Masses (Two Haloes)}\label{subsubsecFateTotalMassesPanel} 


Figure~\ref{figMassPanel} shows the fate total-mass distribution, which is completely different from the distributions at the other stages.  This distribution increases monotonically with mass, except at the very high masses, where a small dip is present.  Reducing $d_{\rm crit}$ and $M_{\rm crit}$ both have the same effect:  to enhance the small-mass end and suppress the large-mass regime.  This effect is stronger for decreasing $M_{\rm crit}$ than for decreasing $d_{\rm crit}$.  Notice that for the non-Massive Sets, the peak of the fate-mass distribution is three orders of magnitude higher than that of the final-mass distribution, while only two orders of magnitude for the Massive Sets.

\subsection{Mass Ratios}\label{subsecMassRatiosPanel}

\begin{figure*}
 \centering
 \includegraphics[width=\hsize]{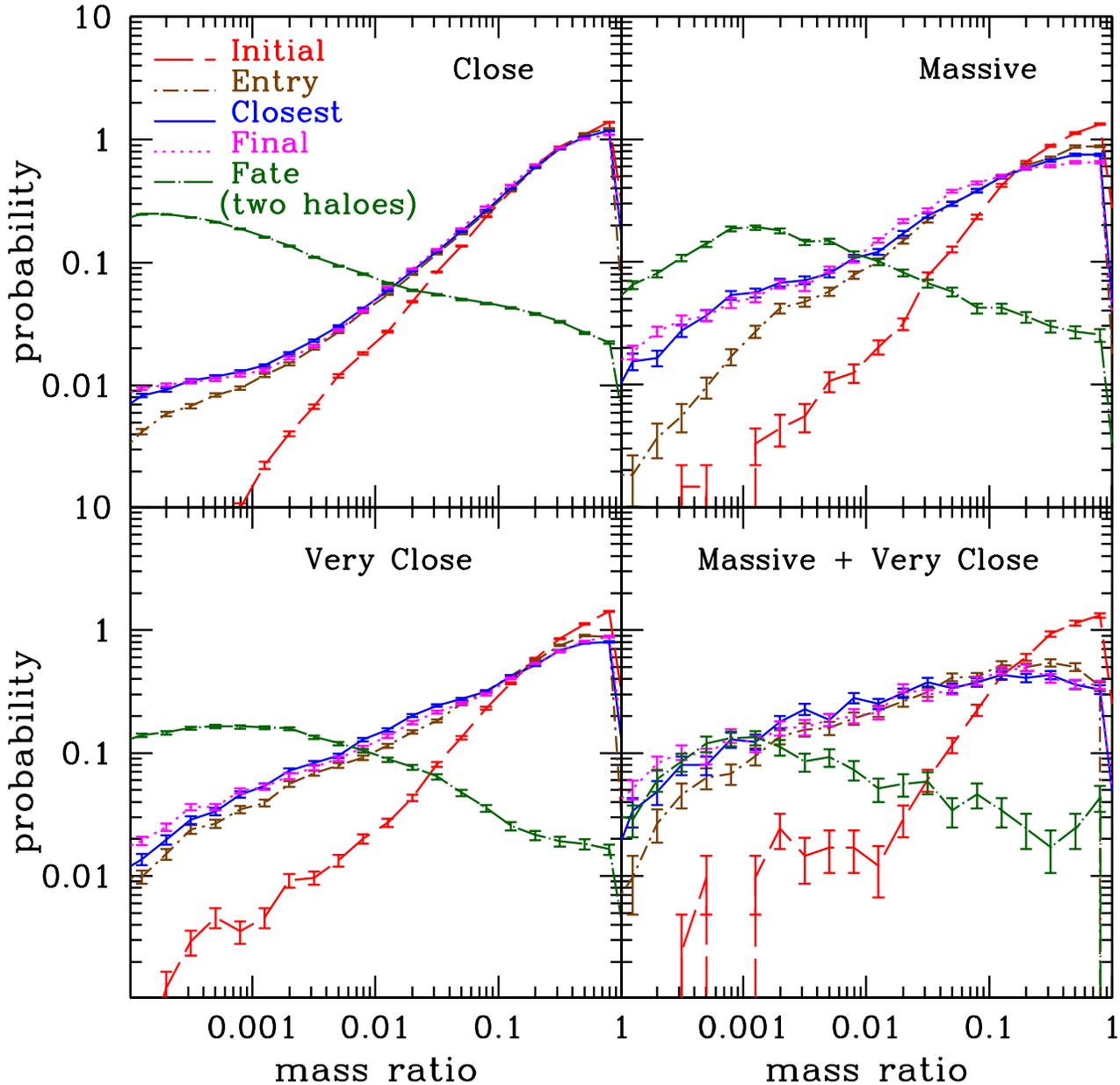}
 \caption{The mass-ratio distribution of halo pairs.   The set of halo pairs presented (see Table~\ref{setsTable}) is indicated in each panel: C (upper-left), VC (lower-left), MC (upper-right) and MVC (lower-right).  The colours and line-styles are indicated in the Figure.  Only pairs that never merger are presented for the fate stage (indicated by the `two haloes' label). For reference, the text uses the following definitions to describe the sets:  Close Sets ($d_{\rm crit}=1$ Mpc $h^{-1}$; upper panels), Very Close Sets ($d_{\rm crit}= 200$ kpc $h^{-1}$; lower panels), non-Massive Sets ($M_{\rm crit}=8.6 \times 10^{10} M_{\odot} h^{-1}$; left-hand panels), and Massive Sets ($M_{\rm crit}=8.6 \times 10^{11} M_{\odot} h^{-1}$; right-hand panels).  See Section~\ref{subsecStages} and \ref{subsubsecSets} respectively for definitions and descriptions of the stages and sets. }
 \label{figMassRatioPanel}
\end{figure*}

Figure~\ref{figMassRatioPanel} shows the distribution of the mass ratios of the haloes in the pairs at the various stages of their evolution.    In this work, the {\it mass ratio} is defined as the ratio of the small to the large mass in a given pair (i.e., this quantity is always $\leq 1$; small values correspond to disparate pairs while large values correspond to nearly-equal pairs).  In particular, the primary versus secondary ranking of the haloes in the pairs need not remain fixed at all times:  at times the haloes may switch roles.  For instance, a halo may begin more massive than its companion, but then the companion could grow quickly for some period and quickly surpass its partner. In these situations, the ranking of the mass ratios are adjusted accordingly.  

The resolution of the simulation, and the criterion that branches must surpass a critical mass threshold, may affect the mass ratio distribution (especially for the initial stage, see Section~\ref{secDiscussion} for a discussion).  Nevertheless, these results are worth presenting because (1) they provide additional clues describing how the pairs in the simulation grow, and (2) they help us understand the effects of changing $d_{\rm crit}$ and $M_{\rm crit}$ more clearly.

The \underline{initial}, \underline{entry}, \underline{closest} and \underline{final} mass-ratio distributions all increase monotonically with mass ratio.  The distribution is sharply peaked initially at very high mass ratios and becomes broader as the pairs evolve (initial $\rightarrow$ entry $\rightarrow$ closest $\rightarrow$ final).  During this process, the low mass-ratio regime gets gradually more populated.  In other words, the masses in pairs tend to become more discrepant with time.  Notice that as this evolution progresses, the distribution is suppressed at mass ratios $>0.1$ and enhanced at mass ratios $<0.1$.   Lastly, the closest and final mass ratios have nearly identical distributions, while the entry distribution is similar to these two, except for a small difference in the low mass-ratio regime.
As in the analysis of the total mass (Section~\ref{subsecTotalMassesPanel}), the \underline{fate} mass-ratio distribution is completely different from the other stages.  For instance, the fate mass ratios are preferentially $<0.1$ (recall too that the sum of the fate-halo masses is preferentially large).

\begin{figure*}
 \centering
 \includegraphics[width=\hsize]{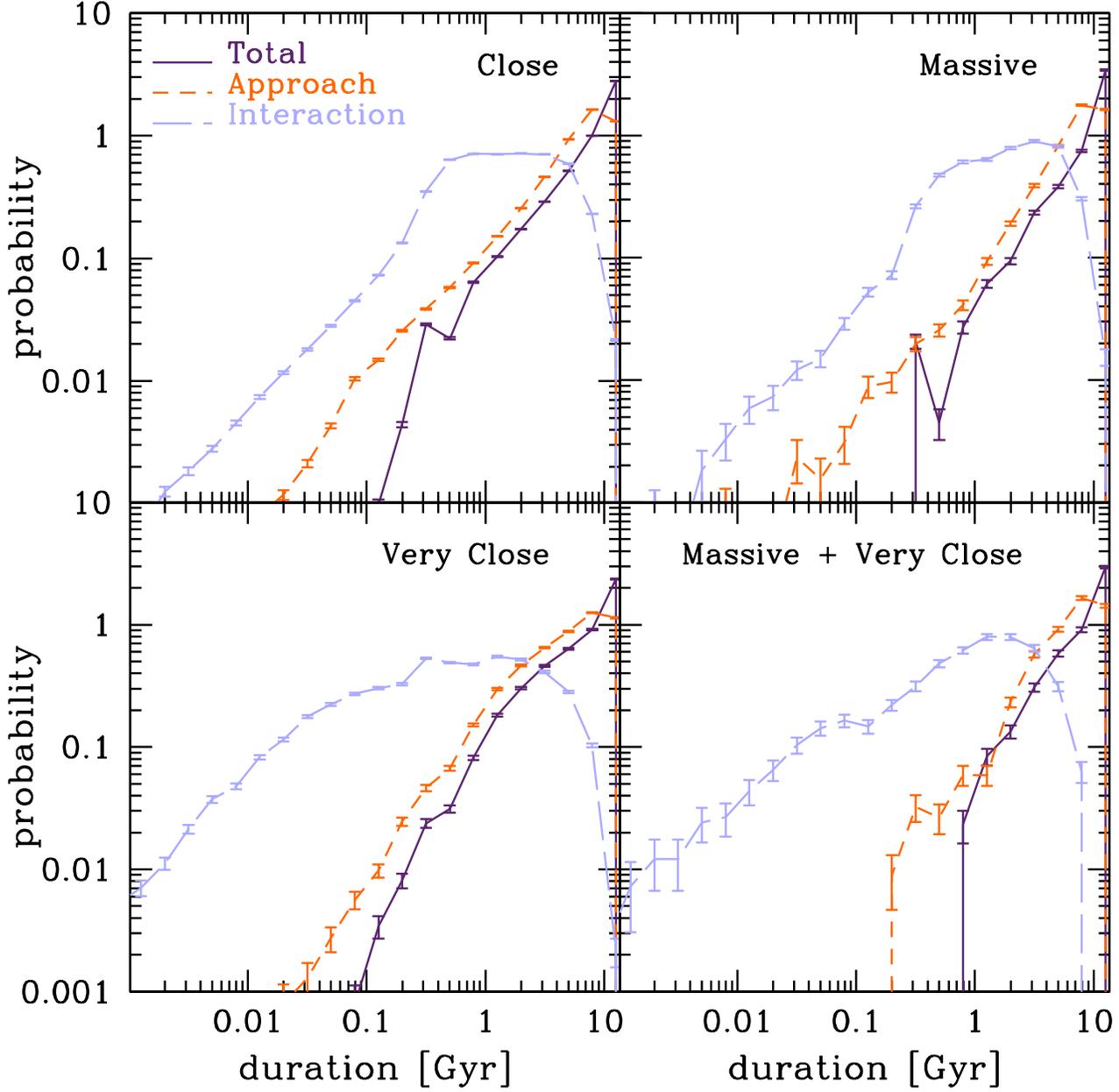}
 \caption{The interaction-duration distribution of halo pairs.   The set of halo pairs presented (see Table~\ref{setsTable}) is indicated in each panel: C (upper-left), VC (lower-left), MC (upper-right) and MVC (lower-right).  The colours and line-styles are indicated in the Figure.  The {\it total} duration refers to the time from the initial until the fate stage.  The {\it approach} time goes from the initial until the entry stage.  The {\it interaction} lasts from the first until the last time the pair is identified within $d_{\rm crit}$.  For reference, the text uses the following definitions to describe the sets:  Close Sets ($d_{\rm crit}=1$ Mpc $h^{-1}$; upper panels), Very Close Sets ($d_{\rm crit}= 200$ kpc $h^{-1}$; lower panels), non-Massive Sets ($M_{\rm crit}=8.6 \times 10^{10} M_{\odot} h^{-1}$; left-hand panels), and Massive Sets ($M_{\rm crit}=8.6 \times 10^{11} M_{\odot} h^{-1}$; right-hand panels).}
 \label{figEvolutionPanel}
\end{figure*}

\subsection{Lifetimes}\label{subsecResultsDuration}

In this Section I estimate the typical duration of close interactions. The following relevant time intervals are considered:

\begin{itemize}

\item {\bf Total Time:}  The interval from the initial to the fate time.  Here, unlike in previous discussionsSection, fate here is not limited to the non-merging pairs.  

\item {\bf Approach Time:}  The interval from the initial to the entry time.  This refers to the approach time from large distances down to the first time the pair is within $d_{\rm crit}$.  Notice that those pairs with $d_{\rm initial} < d_{\rm crit}$ have zero approach time.

\item {\bf Interaction Time:}  The interval from the first to the last time the pair is within $d_{\rm crit}$.  If $d_{\rm initial} < d_{\rm crit}$, the interaction begins at the initial time.  If $d_{\rm initial}>d_{\rm crit}$, it begins at the entry time.  If $d_{\rm fate}<d_{\rm crit}$ (i.e., mergers or sub-critical non-merged pairs), the interaction ends at the fate time.  If $d_{\rm fate}>d_{\rm crit}$, the last time at which the relative separation crosses the $d_{\rm crit}$ threshold is sought.  If this time does not coincide with one of the snapshots in the simulation, a linear interpolation in $\log(1+z)$ is performed.

\end{itemize}

It is important to mention that the approach time depends on the initial time, which itself is strongly determined by the resolution of the simulation.  This issue also affects the estimate of the total time.  Despite this limitation, it is interesting to study the typical timescales experience by close pairs in the Millennium Simulation, and the effects of modifying $d_{\rm crit}$ and $M_{\rm crit}$.  The interaction time is less susceptible to these resolution effects because the mass-criterion adopted in this paper deletes most of the branches that are most prone to go below the mass resolution of the simulation more quickly due to tidal stripping.   Nevertheless, additional improvements using dynamical-friction recipes can always be incorporated in future works (which will most likely increase the duration of the interaction times).  The results presented here serve as a reference for such developments (see Section~\ref{secDiscussion} for additional discussions on these issues).


The three lifetime distributions are shown in Figure~\ref{figEvolutionPanel}.  The \underline{total}-time distribution peaks sharply at long durations, with over 80\% with $t_{\rm total}>5$ Gyr (and over 90\% with such long durations for the Massive Sets).  The \underline{approach}-time distribution resembles the total-time distribution, but with some contributions from shorter durations.  Over 65\% of the pairs have $t_{\rm approach}>5$ Gyr (and over 80\% have such long durations for the Massive Sets).  The \underline{interaction}-time distribution is dominated by durations between approximately $0.2 - 4$ Gyr.  The distribution has a long tail at low-durations (with $t_{\rm interaction} < 0.1$ Gyr), especially for the Very Close Sets ($\sim$20\% for the VC set, and $\sim$12\% for the MVC).




Reducing either $d_{\rm crit}$ or $M_{\rm crit}$ does not affect the distributions significantly. There is a small enhancement in the intermediate time regime (approximately $1-8$ Gyr), but not for the longest-time regimes.  The effect of changing $d_{\rm crit}$ is slightly stronger than that caused by changing $M_{\rm crit}$. Increasing $M_{\rm crit}$ shifts the lower-end of both the total and the approach-time distributions to higher values.  The most obvious change is that
adopting a smaller $d_{\rm crit}$ makes short interactions more common, while choosing a larger $M_{\rm crit}$ makes the approach-times longer.   Lastly, when both modifications are applied, interactions lasting longer than $\sim$8 Gyr are essentially absent.



\section{DISCUSSION}\label{secDiscussion}

This Section discusses the assumptions and provides interpretations for results presented in this paper.  A description of related work is included.  

\subsection{The Assumptions in the Catalogue}\label{subsecDiscussionCatalogue}

Before analysing and intepretting the results of Section~\ref{secResults}, I devote this section to discuss the assumptions adopted in this paper.

The catalogue analysed in this work consists of halo pairs that (1) come within some comoving distance $d_{\rm crit}$ and, (2) in which both haloes reach a mass greater than some threshold value $M_{\rm crit}$ at some point in their history. 

The primary (largest) catalogue is dubbed the `Close Set', and it has $d_{\rm crit}=1$ Mpc $h^{-1}$ and $M_{\rm crit}=8.6 \times 10^{10} M_{\odot} h^{-1}$ (equivalent to 100 simulation particles).  Sub-catalogues are then selected from this set by either decreasing the critical distance to $d_{\rm crit}=200$ kpc $h^{-1}$ or / and by increasing the critical mass to $M_{\rm crit}=8.6 \times 10^{11} M_{\odot} h^{-1}$ (equivalent to $1,000$ simulation particles).  From these combinations, the following additional sets arise:  the Very Close Set, the Massive Close Set and the Massive Very Close Set.  See Table~\ref{setsTable} for a summary.  One of the goals of this work is to quantify the effects of modifying these selection criteria.

The larger critical distance adopted for the primary catalogue ($d_{\rm crit}=1$ Mpc $h^{-1}$) is not as large as it appears because it is in comoving coordinates. 
In these coordinates, this threshold is comparable to the large-separation tail of binary quasars found by \cite{foreman09}.  Indeed, these authors report 38 binaries (out of 85) with comoving separations larger than $200$ kpc $h^{-1}$, and 3 binaries with separations larger than $1$ Mpc $h^{-1}$!  One could argue that such pairs are not truly associated.  However, these authors found their measurements to be enhanced relative to simple extrapolations of the quasar correlation function, signaling that they may indeed be physically influencing each other.

The distance criterion used here is still larger than some of the popular thresholds found in the literature ($\lesssim 50$ physical kpc are commonly used to select pairs with enhanced star formation).  Nevertheless, keeping this exaggerated choice is valuable for a variety of reasons.  First, it facilitates future studies of halo triplets and N-tuplets.  Such investigations can shed light on the evolution of close compact groups -- e.g., like Stephan's quintet \citep{stephan77,renaud10} -- and multiple-quasar systems \citep{djorgovski07,alonso08,liu11a,liu11b,liu11c}. Similarly, the effects of the local environment surrounding the pairs can be tracked \citep[e.g.,][and references therein]{barton07,ellison10}.  Lastly, this large proximity choice will be useful for future studies of sub-Mpc clustering studies of quasars \citep[e.g.,][]{myers07,shen10,bonoli10} and/or luminous red galaxies (LRGs) \citep[e.g.,][]{bell06,masjedi06,masjedi08,wake06,wake08,white07,conroy07,almeida08,brown07,cool08,vandokkum10,tojeiro11}.

Alternatively, one could select pairs by whether they come within a distance below a fixed multiple of the sum of the two virial radii  \citep[e.g.,][]{sinha11}.  However, the appropriate choice of separation for interaction-induced phenomena (e.g., star-formation enhancement) may depend on environment \citep[e.g.,][]{ellison08}.  For the sake of simplicity, a selection based on centre-to-centre distance is preferred at this point.

The mass criterion adopted in the primary catalogue ($M_{\rm crit}=100 \times m_{\rm p}$) is introduced mainly for computational convenience.  Including all possible branches without any mass constraint is extremely cpu-intensive.  Obviously, the problem of interacting dwarf galaxies is very interesting.  However, using the Millennium Simulation for this purpose is not entirely appropriate, since modeling such galaxies with less than a hundred particles is a very crude approximation.  For these systems, using something like the Millennium-II Simulation \citep[with $L_{\rm box}=100$ Mpc $h^{-1}$ and $m_{\rm p}=6.9 \times 10^{6} M_{\odot} h^{-1}$,][]{boylan-kolchin09} is more suitable.  I reserve an extension of this catalogue in that direction for a future project.

\subsection{Interpretation of the Results}\label{subsecDiscussionStages}

The analysis of Section~\ref{secResults} provides a relatively simple picture that incorporates five basic stages:  initial $\rightarrow$ entry $\rightarrow$ closest $\rightarrow$ final $\rightarrow$ fate.  Below I use the wealth of statistical information gathered in Section~\ref{secResults} to make more sense of the details behind this process.  

\subsubsection{The Initial Stage}\label{subsubsecDiscussionInitialStage}

 When both haloes are first identified (above the resolution of the simulation), they are typically found at distances of a few up to 10 Mpc $h^{-1}$, but never beyond 30 Mpc $h^{-1}$ (Figure~\ref{figDistancePanel}).   Figure~\ref{figRedshiftPanel} shows that the pairs usually begin at redshift $\simeq 5 - 7$, with only about $5$\% at $z<1$ (except for the MVC set, which has a $\sim$20\% contribution from this redshift regime -- see the foregoing discussion for a plausible interpretation).  The majority of the total masses in the pairs are initially small:  about 60\% are in the  $(5-10) \times 10^{10} M_{\odot} h^{-1}$ range, $\sim$35\% are in the $(10^{11}-10^{12})\times M_{\odot} h^{-1}$ range, and only $\sim$5\% have masses greater than $10^{12} M_{\odot} h^{-1}$ (Figure~\ref{figMassPanel}).  About 90\% of the initial mass ratios are greater than 0.1 (Figure~\ref{figMassRatioPanel}).

The shape of the initial-distance distribution (Figure~\ref{figDistancePanel}) depends mainly on two competing factors.  As the size of a region around a halo increases, more and more neighbours are within reach.  Eventually, this quantity drops precipitously because the number of branches that participate in the assembly of the final host halo is finite.  One would expect that decreasing the adopted $d_{\rm crit}$ would preferentially select pairs with smaller $d_{\rm initial}$.  Overall, this is not the case, except at small distances:  the Very Close Sets (lower panels of Figure~\ref{figDistancePanel}) exhibit as a small secondary bump at 100 kpc $h^{-1}$ and a trough at 300 kpc $h^{-1}$.  See below for a possible interpretation of this feature.   Indeed, it is reasonable to speculate that with much smaller values of $d_{\rm crit}$, this bump would become more prominent relative to the large-separation peak.  

A large fraction of initial halo pairs are `major' (with mass ratio $>1/10$, Figure~\ref{figMassRatioPanel}).  This is no surprise.  At early times, simulated haloes are expected to have masses slightly higher than the resolution limit ($20 \times m_{\rm p} = 1.72 \times 10^{9} M_{\odot} h^{-1}$), making the corresponding mass ratio of any two haloes $\lesssim 1$.   In reality, one of the haloes often has to `wait' for the other to be born (e.g., branches C and D in Figure~\ref{figTree}).  During this wait period, the primary halo has some time to grow.  However, this growth cannot be too strong, nor too frequent -- otherwise the contribution from pairs with low mass ratios would be more significant.

The nature of the initial stage depends strongly on the capabilities of the simulation.  A simulation with a smaller box would not generate massive enough pairs that begin at such early times, with such large separations.  Likewise, it would not support enough massive final host haloes capable of hosting such massive branches in large numbers.  A simulation with higher resolution, on the other hand, would in principle shift the initial mass distribution to even lower masses and higher redshifts.  However, since the majority of pairs begin at levels not too different from the resolution threshold (which therefore consist, for the most part, of very similar haloes), the initial mass-ratio distribution is not expected to be very susceptible to this.


\subsubsection{The Entry, Closest and Final Stages}\label{subsubsecDiscussionInteractionStages}

Recall that the great majority of pairs fall from rather large distances.  Only $\sim$5\% begin with $d_{\rm initial}<d_{\rm crit}$.  For the rest, it takes a few Gyr to go from $d_{\rm initial} \rightarrow d_{\rm crit}$.   The entry, closest and final separations take place predominantly at low redshifts ($\sim$70-90\% of these events happen at $z<1$, while only $\sim$0.5-5\% occur at $z>3$, Figure~\ref{figRedshiftPanel}).  The total masses at these stages are a few $\times 10^{11} M_{\odot} h^{-1}$ for the non-Massive Sets, and a few $\times 10^{12} M_{\odot} h^{-1}$ for the Massive Sets (Figure~\ref{figMassPanel}).  As expected in hierarchical scenarios of halo growth, increasing $M_{\rm crit}$ shifts the masses to higher values, and redshifts to lower values (left versus right-hand panels of Figures~\ref{figRedshiftPanel} and \ref{figMassPanel} respectively).

Also, as in the initial stage, the mass-ratio distribution at these subsequent stages is dominated by large mass-ratios, but to a smaller degree (Figure~\ref{figMassRatioPanel}).  This is because by the time two haloes come within a critical distance, one of them might have had time to grow (e.g., pair D-E in Figure~\ref{figTree}).  In principle if a simulation with a much higher mass resolution were used -- these distributions would include a less prominent contribution from low-mass ratios simply because there would be more pairs involving two small haloes at all redshifts.  Situations corresponding to massive haloes devouring minuscule ones would not be as important.  This point highlights the fact that mass-ratio statistics depends strongly on the resolution of the simulation.  This limitation is not too problematic for the types of problems I intend to attack with this catalogue (e.g., interacting galaxies, binary quasars, etc.), where cuts in the type of haloes hosting gas-rich galaxies and supermassive black holes must be imposed.

As the evolution of the pair progresses (entry $\rightarrow$ closest $\rightarrow$ final), the corresponding redshift distributions shift towards more recent epochs.  Also, the peak of the total mass distribution shifts slightly towards lower values.  This behaviour is intriguing because one would expect mass to {\it increase} as the pairs (and the Universe) evolve.  However, this shift toward lower masses takes place in the small-mass regime.  Above a few $\times 10^{13} M_{\odot} h^{-1}$ this trend is reversed:  pairs get more massive as their evolution advances.  (Although this effect is minimal for the MVC set:  only a small fraction of low-mass pairs dissolve, and the rest remain in place for the duration of the three stages -- see lower-right panel of Figure~\ref{figMassPanel}.)  This evidence indicates that while massive pairs follow their expected hierarchical-growth, smaller pairs are more susceptible to being stripped off of their mass, since most of them inhabit massive environments like clusters.  A detailed analysis splitting this catalogue as a function of environment will shed light on this effect.  This will be the subject of a future paper.

Additional insight may be gained by looking at the behaviour of the mass-ratio distribution as the evolution continues.  Recall that initially, $\sim$90\% of pairs are identified in the simulation with mass ratios $>1/10$.  By the entry time, this fraction drops to $\sim$80\% for the C set, $\sim70$\% for the MC and VC sets, and $\sim$ 40\% for the MVC set.  The low mass-ratio regime, in turn, gets populated.  In general, the masses of the haloes in pairs become more discrepant with time.   In all cases this can be because, on average,  either the primary haloes grows faster than the second one or the secondary halo dissolves more quickly.  This effect is not expected to be strong because, due to the resolution of the simulation, most pairs begin `major' and remain more or less so.  

From the point of view of redshift, total mass, and mass ratio; the entry, closest and final stages behave rather similarly.  The same cannot be said about the distance distribution, making it a very powerful tool to learn about the differences of these stages.   As in the initial case, distributions are expected to increase with distance (as more neighbours are seen) and drop (because groups have a finite number of branches).  For the most part, the closest-distance distribution only sees the rising portion.  The exception is the MVC set (lower-right panel of Figure~\ref{figDistancePanel}), where the downturn is probed because groups have fewer massive branches satisfying the smaller critical-distance condition.

The distribution of final distances (along with the distribution of never-merging fate distances discussed below) has a couple of interesting peculiarities.  For sub-critical distances, this distribution approximately follows the shape of the closest-distance distribution.   These are the pairs that may, in principle, have the largest chance of merging.  However, there is also a large fraction of pairs ending up with $d_{\rm final}>d_{\rm crit}$.  This is expected because pairs are selected by whether or not they came within some critical separation, instead of by following imminent mergers back in time.  

Surprisingly, the fraction of super-critical final distances {\it increases} when $d_{\rm crit}$ is reduced (from $\sim$45\% $\rightarrow$ $\sim$57\% for the non-Massive Sets and from $\sim$39\% $\rightarrow$ $\sim$45\% for the Massive Sets -- see portions to the right of the vertical line in Figure~\ref{figDistancePanel}).  A possible interpretation is as follows:  as the adopted critical distance increases, the number of close neighbours a given halo can have also increases.  While the number of massive neighbours increases, the number of less massive haloes increases more substantially.  In other words, the importance of having massive haloes in the vicinity diminishes (relative to the less-massive neighbours) with increasing $d_{\rm crit}$.  Conversely, having a small $d_{\rm crit}$ means that the effects of third-party massive haloes in the neighbourhood (probably the main halo within the same group) become more dominant.  One such effect is to split halo pairs:  while two haloes may accompany each other for extended periods, one of them may be absorbed by a third massive halo before they have the opportunity to merge.  Another effect is to momentarily separate the pairs into very elongated orbits at the very last moments.  Alternatively, these orbits might be elongated simply because when the distance between two haloes is comparable to their virial radii, their orbits tend to be rather eccentric \citep{zentner05}.  In either case, at the final stage, a large fraction of halo pairs are observed with separations much larger than $d_{\rm crit}$.  Indeed, it would be quite interesting to be able to track these last stages with more detail, using a simulation with much higher resolution. 

On a side note, if the importance of massive neighbours increases with decreasing $d_{\rm crit}$, a larger contribution from low mass-ratio pairs should be expected .  Figure~\ref{figMassRatioPanel} confirms this enhancement for all the stages mentioned here in (upper versus lower panels).  Nevertheless, this effect does not change the fact that the great majority of mass ratios remain more or less `major' at all times.

Finally, I must emphasize that the notions of closest and final stages should not be taken too literally.  If two haloes merge, the closest and final distances they attain are strictly zero.  My motivation to introduce these stages is not in this strict sense.  Instead, my intention is to give an approximate picture (albeit snapshot-limited) of how the pairs evolve in the simulation.

\subsubsection{The Fate Stage}\label{subsubsecDiscussionFateStage}

It is important to keep track of the ultimate fate of the pair, even though this stage is not formally part of the evolution (although an infinitesimal of time before it actually is).  As expected, reducing $d_{\rm crit}$ promotes the contribution of mergers, consequently reducing the fraction of never-merging pairs (from $\sim$53\% $\rightarrow$ $\sim$42\% for the non-Massive Sets, and from $\sim$39\% $\rightarrow$ $\sim$27\% for the Massive Sets).  

For non-merging pairs, the fate-redshift distribution is strongly shifted towards recent epochs (fate-pairs with $z<0.5$ amount to $\sim$50\% for the non-Massive Sets, and $\sim$60\% for the Massive Sets -- see Figure~\ref{figRedshiftPanel}).  At these late epochs, massive groups and clusters become more prominent.   These dense environments may be responsible for preventing the mergers of otherwise close pairs of haloes.

Disruption of pairs can also be mitigated by adopting a larger $M_{\rm crit}$.  This is because pairs of massive haloes are less prone to be disrupted by third-pary haloes.  In principle, only super-massive neighbours could actually achieve these splittings, but such extreme objects are not very common.  

So far I have suggested that some pairs of haloes never merge because one or both haloes participating in the interaction are absorbed by a third party.  In practice, none of the fate pairs get absorbed simultaneously by two distinct haloes.  Instead, it is common to have only one of the two haloes in the pair be eaten by a third object, leaving the other one wandering alone. (Moreover, in many cases the left-over halo is also absorbed by the same third object, most likely the central halo, but at a much later time.)   This picture is supported by the following pieces of evidence.   First, $\sim$80\% of the non-merging haloes end up with fate total masses $>10^{13} M_{\odot} h^{-1}$.   Second, $\sim$90\% of this subset have fate mass-ratios $<1/100$, supporting the interpretation that single, very massive third-party haloes are responsible for ripping the pairs apart.  

The fate distance distribution, just like the final distance distribution discussed before, contains valuable spatial information.  First of all, it is useful to compare the fraction of fate non-merging pairs with the fraction of super-critical final pairs.  For the C, VC, MC, and MVC sets, recall that the super-critical final fractions are $\sim$45\%, 55\%, 40\% and 45\%; while, on the other hand, the non-merging fate fractions (quoted above) are $\sim$53, 42\%, 39\% and 27\%.  This demonstrates that the set of super-critical final pairs do {\it not} fully determine the set of never-merging fate pairs (read below for a possible interpretation).  

Nevertheless, if we focus on the super-critical fate regime instead, the corresponding fractions (relative to the general halo-pair population) are $\sim$22\%, 35\%, 14\% and 27\%.  This demonstrates two things.  First, that the fraction of super-critical fate pairs is enhanced with reduced $d_{\rm crit}$, displaying the same behaviour as their super-critical final counterparts.  Second, that only a small fraction of the super-critical final pairs never merge.  A significant fraction instead do {\it return} to sub-critical separations and merge eventually.  This restriction of the snapshot-limited final stage can be ameliorated by including velocity information across the last few snapshots.  

Likewise, the selection of close pairs by their closest separation can be limited:  pairs with separation $>d_{\rm crit}$ at two consecutive snapshots, but with separation $<d_{\rm crit}$ between these two snapshots, are automatically missed by the selection process.  However, these instances are not expected to happen too frequently, and such pairs are not likely to penetrate close enough towards each other and have enough time to produce significant tidal damage.  In any case, improvements like these can be incorporated in future, more detailed analyses. 

It is interesting to concentrate also on the fraction of super-critical fate pairs relative to the sample of never-merging pairs (as opposed to the general population).  The fractions are $\sim$40\%, 80\%, 35\%, 78\%.  This finding, along with the behaviour of the total-mass and mass-ratio distributions at the fate stage, support the idea that reducing $d_{\rm crit}$ augments the influence of massive neighbours on pairs that never merge.   The importance of super-critical pairs with small $d_{\rm crit}$ is more considerable at the fate stage than at the final stage.   This is because the latter cases are a special subset of the former: they are the ones that are split apart more powerfully.

This analysis suggest that single massive haloes (perhaps the main branch, but other massive secondary branches as well) enhance separations at the final stage and ultimately help split close pairs.  The choice of adopted $d_{\rm crit}$ is very important in this regard:  small values of critical distance give more weight to the influence of these third-party massive objects.   At this point I speculate that further reducing $d_{\rm crit}$ will certainly mitigate the fraction of never-merging pairs, but those that never merge will be separated at even more violent levels.  On the other hand, while it is true that increasing $M_{\rm crit}$ enhances mergers, the fraction of super-critical fate pairs is left largely untouched by this modification because there are not large numbers of supermassive neighbouring haloes to rip them apart at these impressive levels.

\subsubsection{The Lifetimes}\label{subsecDiscussionLifetime}



The lifetime of a halo pair goes from the $t_{\rm initial}$ to $t_{\rm fate}$.  In general, the total duration of the pairs is very long, with over $\sim$97\% of the pairs lasting more than 1 Gyr.   Notice that increasing $M_{\rm crit}$ increases the fraction of pairs that live more that 5 Gyr (from $\sim$86\% $\rightarrow$ $\sim$91\% for the Close Sets, and from $\sim$78\% $\rightarrow$ $\sim$89\% for the Very Close Sets -- left versus right-hand panels in Figure~\ref{figEvolutionPanel}).  Reducing $d_{\rm crit}$, on the other hand, makes the total lifetimes shorter (upper versus lower panels in the Figure).  It is possible that long total times are mitigated because adopting smaller critical distances tends to pre-select pairs that have a higher probability of merging.

The approach period, which goes from $t_{\rm initial}$ to $t_{\rm entry}$, takes up a dominant fraction of the total time.  In many ways, $t_{\rm approach}$ exhibits similar trends to $t_{\rm total}$.  Over $\sim$95\% of pairs have approach times greater than 1 Gyr. Increasing $M_{\rm crit}$ increases the fraction of pairs with approach time greater than 5 Gyr (from $\sim$77\% $\rightarrow$ $\sim$84\% for the Close Sets, and from $\sim$66\% to $\sim$80\% for the Very Close Sets). As is the case for the total time, reducing $d_{\rm crit}$ also shrinks the approach time.  However, mergers do not happen at the end of the approach period.  Nevertheless, adopting a small critical distance selects pairs of haloes that are more prone to merge eventually, even if this happens long after the end of the approaching phase.

As mentioned before, the total and approach times depend largely on the capabilities of the simulation.  A simulation with either a larger box or higher resolution would in principle be capable of having even longer times (limited, of course, by the age of the Universe).  A larger box allows more massive haloes with massive branches that are born earlier.  Despite these limitations, it is important to report the capabilities of the Millennium Simulation in regards to the evolution of close halo pairs.  Higher resolution also allows all haloes to be born earlier, which can in turn be tracked down for longer periods, and more accurately.  Even at late stages, small dissolving haloes in dense environments could be followed down for longer times (most likely making the interaction times longer) if a simulation with higher resolution were used \citep[e.g.,][]{boylan-kolchin09}.  Alternatively, one can also apply dynamical-time recipes to track these structures \citep{binney87}.

Simulations also limit the short-time regime.  In particular, given that $t_{\rm total}$ is defined as the difference of two epochs associated with two simulation snapshots (i.e. $t_{\rm total}\equiv t_{\rm fate}-t_{\rm initial}$) results in an artificial lower bound governed by the largest time snapshot of the simulation (for the Millennium Run, this corresponds to $\Delta t=0.0122$ Gyr).  


The interaction time goes from the entry-time up until the last time the pair is within $d_{\rm crit}$.  For merging and sub-critical fate pairs, this ends at $t_{\rm fate}$; otherwise it ends before.  Unlike the total and approach times, the distribution of interaction times has a noticeable short-time tail.  This tail is more evident for the smaller choice of $d_{\rm crit}$, where the prevalence of mergers suppresses the duration of these phases.  For instance, the fraction of pairs with interaction times {\it shorter} than 1 Gyr increases from $\sim$40\% $\rightarrow$ $\sim$60\% for the non-Massive Sets, and from $\sim$30\% $\rightarrow$ $\sim$50\% for the Massive Sets.  

Using a large $M_{\rm crit}$ inhibits the fraction of pairs that interact for long periods because these pairs are more likely to spend most of their lives in the approach phase (see Section~\ref{subsecResultsDuration}).  In all cases, the contribution of pairs with durations longer than 8 Gyr is negligible.

Strictly speaking, the time spent within $d_{\rm crit}$ may be shorter than the $t_{\rm interaction}$ reported in this paper because, in many cases, the secondary halo may oscillate out and back inside the threshold on multiple occasions.  Such instances are probably more important for subhaloes orbiting a primary massive (perhaps central) halo \citep[e.g., ][]{benson05,khochfar06,angulo09,wetzel11}.   This level of detail is currently beyond the scope of this work, and can always be incorporated in more refined versions of the catalogue.

\subsection{Relation to Other Theoretical Work}\label{subsecComparison}

The nature of the evolution of pairs of haloes (or galaxies) from a cosmologically point of view remains relatively unexplored.  This section is devoted to discussing those few works that have attempted to rise beyond pre-prepared numerical experiments of a single nearly-isolated pair of galaxies.  Here I simply discuss how their methods and assumptions compare to mine.  However, a direct comparison of results is not possible at this point because of the different set of goals of those papers and the present work.

\cite{tissera02} are among the first research groups to attempt to simulate interactions from a cosmological perspective.   Smooth particle hydrodynamical (SPH) simulations are employed to study starbursts in merging galaxies.  Their results suggest a double-triggering scenario where the first starburst is provoked early on in the interaction.  Unfortunately, to attain such detail, they use a very small volume, $L_{\rm box}=5$ Mpc $h^{-1}$ (with $h=0.5$) on the side, and only $64^3$ particles.   Follow-up work by \cite{perez06} uses a volume with $L_{\rm box}=10$ Mpc $h^{-1}$ and $\sim 80^3$ particles.  The scope of this approach can be thought to be `in between' the single-pair simulations and the present work.

The approach of \cite{sinha11} is very similar to mine, in the sense that they also focus on dark-matter only simulations \citep[unlike][who include gas dynamics]{tissera02,perez06}.  These authors propose a notion called the `halo interaction network', and suggest that fly-bys at high redshift ($z\gtrsim10$) are more frequent than mergers.  However, their simulation has $L_{\rm box}=50$ Mpc $h^{-1}$ on the side, {\it only one thousandth} of the volume in the Millennium Simulation.  Their box size  is comparable to the initial separations of pairs presented here (Figure~\ref{figDistancePanel}).   It is not clear how representative their particular patch (or the merger histories within) actually is; or if they can produce enough massive background haloes where the majority of the halo pairs happen to exist.  See Appendix~A of \cite{wetzel10} for a discussion on the impact of using a small simulation box on the clustering of subhaloes.

\cite{kitzbichler08} also employ the Millennium Simulation, along with the semi-analytic galaxy formation model of \cite{croton06} and \cite{delucia07}.  Their mock catalogue is extracted from a narrow light-cone with a field of view of $10\times1.4$ deg$^{2}$.  Their goal is very specific: to calibrate the time-scales connecting close galaxy pairs and their subsequent mergers.  Given the nature of their objective, they are required to refine the stages immediately prior to the mergers.   To do this, they include additional dynamical friction times in their estimates \citep{binney87,boylan-kolchin08}.  Improvements like this can certainly be implemented in the estimates shown in Figure~\ref{figEvolutionPanel}.

On a similar vein, \cite{berrier06} use a simulation with $L_{\rm box}=120$ Mpc $h^{-1}$ on the side.   Following halo mergers, these authors employ the analytic substructure model of \cite{zentner05} to track down subhaloes as they sink in their host groups.  Their goal is also very concrete:  to evaluate the number of close companions per galaxy.  They argue that instead of using close halo pairs to predict the halo merger rate, the number of companions is a more suitable quantity to test various galaxy formation scenarios.  This research group then takes advantage of that approach to search simulated galaxy-pairs in isolation \citep{barton07} and investigate star formation enhancements provoked by interactions.  The types of selection criteria applied in these papers will be very valuable in future studies of halo pairs in different environments, and in more realistic applications of my catalogue.  

\cite{wetzel08} also investigate how reliable close halo pairs are as proxies of mergers.  However, they only concentrate on mergers of cluster-sized haloes (with masses $>5 \times 10^{13} M_{\odot} h^{-1}$) at low redshifts ($z<1$).  To do this, they track pairs of `halo objects' (i.e., the equivalent of an isolated halo in this paper), at the expense of ignoring subhaloes altogether.  In a way, that work is complementary to mine:  they manage to relate pairs that end up in different final groups but dismiss pairs of subhaloes within a given group.  Pairing up {\it all} possible haloes and subhaloes at a given distance in a simulation, and tracking their evolution to test whether they merge or not, is a very computationally cumbersome problem.  A possible way to extend my approach in that direction is to compare branches that end up within a small bundle of groups in the local universe.  However, at this point it is not clear how large the volume containing these groups should be.   These difficulties illustrate how challenging the general study of close pairs as proxies of mergers can actually be.

On a slightly different subject, \cite{wetzel08} also find that while rich environments have more halo pairs, mergers of such pairs are more likely to take place in less-dense environment.  With the analyses presented in this work I argue that mergers are inhibited when the pairs involved happen to live in crowded environments.  While it is true that this is reassuring, a direct comparison between the two works is impossible at the moment because their environmental scale is $\simeq 17$ Mpc $h^{-1}$, while mine is restricted to regions that end up as single host haloes.

\section{SUMMARY}\label{secSummary}

This paper describes a statistical analysis of a very large catalogue of co-evolving halo pairs I constructed from the merger history trees in the publicly available Millennium Simulation Database \citep{springel05}.    

The catalogue itself is comprised of over half a million pairs, selected by whether they come within a given critical separation $d_{\rm crit}$ (Figure~\ref{figTree}).  Only haloes that attain a mass larger than $M_{\rm crit}$ at some point in their history are chosen (Figure~\ref{figThreshold}).  Sub-catalogues are constructed to study the effects of modifying these criteria. The adopted values are $d_{\rm crit}=1$ and $0.2$ Mpc $h^{-1}$;  and $M_{\rm crit}=8.6 \times 10^{10}$ and $10^{11} M_{\odot} h^{-1}$ (corresponding to 100 and 1,000 simulation particles respectively).  See Table~\ref{setsTable} for a summary of four sets arising from these combinations.  

In order to shed light on the evolution of these pairs, I establish a simple picture consisting of five basic stages:  {\it initial} $\rightarrow$ {\it entry} $\rightarrow$ {\it closest} $\rightarrow$ {\it final} $\rightarrow$ {\it fate} (see Section~\ref{subsecStages}).  Also, the following relevant time-scales are considered: the {\it total}, {\it approach} and {\it interaction} times (see Section~\ref{subsecResultsDuration}).

Below I highlight some of the most interesting points we learn from this picture:

\begin{itemize}

\item The majority of pairs begin at $z\gtrsim4$, with separations $\simeq10$ Mpc $h^{-1}$.  The initial distance distribution is largely insensitive to $d_{\rm crit}$, except for the subcritical pairs (with $d_{\rm initial}<d_{\rm crit}$, only $\sim$5\% of them). 

\item The distribution of the sum of the initial masses peaks at a few $\times 10^{10} M_{\odot} h^{-1}$, regardless of $M_{\rm crit}$.  Initially, the two haloes in a given pair tend to be similar, with mass ratios typically greater than $1/10$.

\item The small portion of sub-critical initial pairs is likely to correspond to the surviving sub-structures in cluster-sized host haloes.  This is supported by the end-of-the-tail effects at low redshifts, high total-masses and low mass-ratios.

\item  As the evolution advances (entry $\rightarrow$ closest $\rightarrow$ final), total masses increase steadily if they are $\gtrsim 10^{13} M_{\odot} h^{-1}$, otherwise, they tend to decrease.  This particular mass scale marks the range where tidal stripping becomes relevant.  Mass ratios tend to decrease, making the discrepancy between the two haloes in the pair more evident.

\item Adopting a larger $M_{\rm crit}$ shifts the mass distribution in the interacting stages to higher values:  from $10^{11} \rightarrow 10^{12} M_{\odot} h^{-1}$.   This also suppresses the high-redshift regime (with $z \gtrsim 4$) significantly (for the stages after the entry time). 

\item The final distance distribution peaks at $d_{\rm crit}$.  Around $40-50\%$ of the pairs end up with super-critical separations (with $d_{\rm final}>d_{\rm crit}$).   Some of these pairs may be on their way to becoming permanently separated, while others may be temporarily on elongated orbits, bound to return together and merge eventually. 

\item $\sim$53\% of the pairs in the primary catalogue never merge (these include pairs in which one of the haloes is absorbed by a third halo or pairs that do not merge by $z=0$, but may merge in the future).   Reducing $d_{\rm crit}$ turns this number into $\sim$42\%; while enhancing $M_{\rm crit}$ turns it into $\sim$39\%.  With both effects, one gets $\sim$27\%.  Over half of these non-merging pairs take place at $z<0.5$.

\end{itemize}

In general, merging pairs are more likely to be selected when the critical distance threshold is reduced.  However, the relative fraction of never-merging pairs with super-critical fate separations is enhanced with this reduction.  The same happens with the super-critical final separations.  Squeezing the proximity criterion facilitates the selection of mergers, but also selects never-merging pairs that get ripped apart more violently.  This hypothesis is supported by the fate total-masses and mass-ratios, which suggest a picture in which massive third haloes in the vicinity are responsible for preventing mergers.  Indeed, it would be interesting to see if this effect remains with a simulation capable of following subhaloes with much higher mass resolution.  Overall, a higher $M_{\rm crit}$ also induces mergers.  This is because super-massive third-party haloes in the vicinity capable of splitting the pairs are less common.  However, the violent super-critical effects are less important with this choice.

The second aim of this paper is to provide rough estimates on the typical lifetimes of the pairs, and the duration of their interactions.   Some of the main findings are:

\begin{itemize}

\item On average, halo pairs last a few Gyr, spending most of this time in their approach phase.  Interactions (from the first to the last time the haloes are within $d_{\rm crit}$) last from a few hundred Myr to a few Gyr.  Interactions are never longer than $\sim$8 Gyr.  
\item Adopting a smaller $d_{\rm crit}$ promotes the fraction of mergers, shortening the typical interaction time as a consequence.  Approach and total times are also reduced, but this effect is less noticeable.

\item Adopting a larger $M_{\rm crit}$ produces longer lifetimes.  This is because massive haloes are usually born earlier.  However, this increase in duration happens mostly in the approaching stages (between the initial and entry times).  

\end{itemize}

In the discussion I mention that the nature of the initial stage depends strongly on the resolution of the simulation (which in turn affects the statistics of the approach and total times).  Nevertheless, it is useful to have an idea of the capabilities of the Millennium Simulation in regards to close pairs at the highest-possible redshifts.  The mass-ratio distribution is also susceptible to the low-mass cut in the simulation.  However, the types of applications the catalogue in this paper is ultimately intended for will also include necessary cuts in the types of haloes used.  Lastly, resolution also affects subhaloes in the late stages of interaction.  Higher resolution or the adoption of dynamical-friction recipes should, in principle, provide better estimates in this regime.  In particular, interaction times are expected to be longer than those reported here.

This paper marks the foundation of a series of papers dedicated to the analysis of this huge halo-pair catalogue.  For the sake of space, several studies must be postponed for forthcoming papers.  For instance, the basic quantities discussed here (separation, redshift, total mass and mass ratio) constitute a four-dimensional parameter space, whose exploration is prohibitively unmanageable to be addressed in this exploratory paper.  The same is true for the environment or kinematic quantities like relative velocities and spins. These studies are reserved for forthcoming papers.

Ultimately, this catalogue will be the basis of future investigations of galactic interactions and double (and multiple) quasar systems from a cosmological perspective.  Such studies will require more detailed analyses, including the incorporation of semi-analytic recipes and the handling of selection effects.  For this reason, preliminary studies like the present work are a vital step.  The results in this introductory paper will be essential pieces to guide this effort.

\section*{acknowledgments}
This work is supported in part by SISSA Young Scientist Research Grant ASTR. 639 and by CONACyT-Mexico.    I wish to thank Joanne Cohn, Andrew Benson, Carlo Baccigalupi, Ravi Sheth, Gabriella de Lucia and Monica Colpi for carefully reading an earlier version of this manuscript and for their insightful comments and encouragement.  I also thank Francesco Shankar for discussion on the subject, and both Federica Bianco and Carlo Giocoli for their help with Supermongo.  I thank the anonymous reviewer for her/his useful and timely report, which truly helped improve the quality of this paper. I acknowledge Gerard Lemson and Gabriella de Lucia for their endless patience with my questions regarding the Millennium Simulation.  Also, I thank Viridiana Moreno for her help with running and baby-sitting the code that made this catalogue a reality.  Lastly, I express my gratitude to Luigi Danese, Paolo Salucci and Carlo Baccigalupi for making SISSA such an awesome place to work.  The Millennium Simulation databases used in this paper and the web application providing online access to them were constructed as part of the activities of the German Astrophysical Virtual Observatory. Extensive use was made of the Millennium Database query instructions \citep{lemson06} and the data visualisation package \textsc{topcat} \citep{taylor05} in the early stages of this project.



\bibliographystyle{mn2e}

\label{lastpage}
\end{document}